\documentclass[aps,pra,reprint,superscriptaddress,showpacs]{revtex4-1}

\usepackage{graphicx}% Include figure files
\usepackage{dcolumn}% Align table columns on decimal point
\usepackage{bm}% bold math
\usepackage{color}
\usepackage{scrextend}		
\usepackage{url}
\usepackage{epstopdf}

\begin{document}

\title{Investigation of a possible electronic phase separation in the magnetic semiconductors Ga\textsubscript{1-x}Mn\textsubscript{x}As and Ga\textsubscript{1-x}Mn\textsubscript{x}P by means of fluctuation spectroscopy}

\author{Martin Lonsky}
\email{lonsky@physik.uni-frankfurt.de}
\author{Jan Teschabai-Oglu}
\affiliation{Institute of Physics, Goethe-University Frankfurt, 60438 Frankfurt (M), Germany}
\author{Klaus Pierz}
\author{Sibylle Sievers}
\author{Hans Werner Schumacher}
\affiliation{Physikalisch-Technische Bundesanstalt, Bundesallee 100, 38116 Braunschweig, Germany}
\author{Ye Yuan}
\author{Roman B\"{o}ttger}
\author{Shengqiang Zhou}
\affiliation{Helmholtz-Zentrum Dresden-Rossendorf, Institute of Ion Beam Physics and
Materials Research, Bautzner Landstrasse 400, 01328 Dresden, Germany}
\author{Jens M\"{u}ller}
\affiliation{Institute of Physics, Goethe-University Frankfurt, 60438 Frankfurt (M), Germany}

\date{\today}

\begin{abstract}
We present systematic temperature-dependent resistance noise measurements on a series of ferromagnetic Ga$_{1-x}$Mn$_{x}$As epitaxial thin films covering a large parameter space in terms of the Mn content $x$ and other variations regarding sample fabrication. We infer that the electronic noise is dominated by switching processes related to impurities in the entire temperature range. While metallic compounds with $x>2$ \% do not exhibit any significant change in the low-frequency resistance noise around the Curie temperature $T_\mathrm{C}$, we find indications for an electronic phase separation in films with $x<2$ \% in the vicinity of $T_\mathrm{C}$, manifesting itself in a maximum in the noise power spectral density. These results are compared with noise measurements on an insulating Ga$_{1-x}$Mn$_{x}$P reference sample, for which the evidence for an electronic phase separation is even stronger and a possible percolation of bound magnetic polarons is discussed. Another aspect addressed in this work is the effect of ion-irradiation induced disorder on the electronic properties of Ga$_{1-x}$Mn$_{x}$As films and, in particular, whether any electronic inhomogeneities can be observed in this case. Finally, we put our findings into the context of the ongoing debate on the electronic structure and the development of spontaneous magnetization in these materials.            
\end{abstract}

%%%%%%%%%%%%%%%%%%%%%%%%%%%%%%%%%%%%%%%%%%%%%%%%%%%%%%%%%%%%%%%%%%%%%%%%%%
\pacs{Valid PACS appear here}% PACS, the Physics and Astronomy

\maketitle

\section{Introduction} \label{intro}
Diluted magnetic semiconductors (DMS) combine the benefits of semiconducting and magnetic materials and hence are most promising candidates for future spintronics applications \cite{Dietl2014, Dietl2010}. In particular, III-Mn-V DMS, as for instance the archetypal compound Ga$_{1-x}$Mn$_{x}$As, have been the subject of intensive research in the last two decades, motivated by both fundamental and technological interests \cite{Ohno1998}.  
In Ga$_{1-x}$Mn$_{x}$As, as well as in the related compound Ga$_{1-x}$Mn$_{x}$P, a few atomic percent of the nonmagnetic gallium host sublattice atoms are replaced with magnetic manganese atoms acting as acceptors. Both materials exhibit a long-range magnetic order of substitutional Mn ions mediated by holes \cite{Jungwirth2006, Jungwirth2007, Scarpulla2005, Winkler2011}. Crystalline defects---in particular Mn interstitials (Mn\textsubscript{I}), which act as double donors and thereby compensate the hole doping, as well as As antisites (As\textsubscript{Ga})---play a substantial role for both the strongly interrelated electronic and magnetic properties of DMS. Therefore, the preparation of high-quality samples is a delicate procedure and constitutes an obstacle to further enhancement of the Curie temperature $T_\mathrm{C}$ beyond $190\,$K for Ga$_{1-x}$Mn$_{x}$As \cite{Chen2009} and $65\,$K for Ga$_{1-x}$Mn$_{x}$P \cite{Farshchi2006}. 
Moreover, the effects of disorder \cite{Berciu2001} and carrier-carrier interactions \cite{Richardella2010} make the theoretical description difficult \cite{Sato2010}, and there is still no consensus on the development of spontaneous magnetization in these materials.  
In literature, a broad spectrum of sometimes opposing theoretical approaches exists \cite{Samarth2012}, ranging from the assumption of free charge carriers \cite{Koenig2003, Dietl2001, Dietl2000, Koenig2000} to the opposite case of strongly localized carriers \cite{Dobrowolska2012, Bhatt2002, Berciu2001}. While the first picture assumes the Fermi energy to lie within a merged band consisting of the valence band and a strongly broadened impurity band, the second model proposes the existence of a separate impurity band.
The latter, so-called impurity-band model is expected to be valid for weakly-doped samples. In this context, Kaminski and Das Sarma \cite{Kaminski2002, Kaminski2003} have developed an analytic polaron percolation theory for DMS ferromagnetism in the limit of strong charge carrier localization and for an inhomogeneous spatial distribution of magnetic impurities. In this model, the spins of localized charge carriers can polarize the surrounding magnetic impurities, leading to the emergence of bound magnetic polarons (BMP), which grow in size for decreasing temperatures and at low enough temperatures overlap until an infinite cluster is formed and spontaneous magnetization occurs \cite{Kumar2013}.
Another conceivable scenario is that the p-d Zener model, which is equivalent to a weak-coupling RKKY picture and believed to be appropriate for more metallic systems with free charge carriers, can also be applied on the insulator side of the metal-insulator transition (MIT) of Ga$_{1-x}$Mn$_{x}$As. In this case, the hole localization length remains much greater than the average distance between the acceptors \cite{Ohno2008, Dietl2006}. Large mesoscopic fluctuations in the local value of the density of states near the MIT are expected to lead to a nano-scale phase separation into ferromagnetic and paramagnetic regions below $T_\mathrm{C}$, whereby the paramagnetic (hole-poor) regions can persist down to low temperatures, coexisting with ferromagnetic (hole-rich) bubbles. \newline
Motivated by the still ongoing debate about the particular mechanisms of ferromagnetism and the crucial role of point defects, the present study focuses on a systematic investigation of low-frequency electronic transport properties on a series of Ga$_{1-x}$Mn$_{x}$As samples with different Mn contents and growth parameters. We apply fluctuation spectroscopy, a method which is sensitive to electric inhomogeneities on the nano- and micrometer scale and has been proven to be a versatile tool for identifying electronic phase separation and magnetically driven percolation, as observed, e.g., for perovskite manganites \cite{Podzorov2000} and more recently for the semimetallic ferromagnet EuB\textsubscript{6} \cite{Das2012}. Here, we aim to compare the noise behavior of metallic Ga$_{1-x}$Mn$_{x}$As films ($x>2$ \%) with rather insulating ($x<2$ \%) samples in order to gain further insight into possible percolative transitions and electronic phase separation. Moreover, we investigate another compound, Ga$_{1-x}$Mn$_{x}$P, an even better candidate for the observation of a percolative transition. Despite its chemical similarity with Ga$_{1-x}$Mn$_{x}$As, Ga$_{1-x}$Mn$_{x}$P has a Mn acceptor level lying four times deeper within the band gap, i.e.\ about $0.4\,$eV above the valence band edge \cite{Clerjaud1985}. Therefore, it is obvious that the charge carriers are of a much more localized nature than in Ga$_{1-x}$Mn$_{x}$As. Nevertheless, hole-mediated ferromagnetism has also been demonstrated in Ga$_{1-x}$Mn$_{x}$P \cite{Farshchi2006}. In addition, Ga$_{1-x}$Mn$_{x}$P is very similar to Ga$_{1-x}$Mn$_{x}$As concerning the magnetic anisotropy, spin-polarization and the scaling of $T_\mathrm{C}$ as a function of the Mn concentration \cite{Zhou2015}. A comparison of weakly doped Ga$_{1-x}$Mn$_{x}$As with a Ga$_{1-x}$Mn$_{x}$P sample ($x=3.5$ \%) may provide valuable information about possible commonalities and differences with regard to their noise characteristics and possible electronic inhomogeneities. Apart from varying the Mn content of the above mentioned compounds, an alternative approach for the control of magnetic and electronic properties is the irradiation with He ions, leading to the introduction of deep traps into the system and thereby to an increasing disorder. In this work, we therefore also address the question of which consequences ion irradiation has for the low-frequency resistance noise characteristics and whether signatures of an electronic phase separation or introduced defects can be observed for irradiated samples. It should be noted that fluctuation spectroscopy in general is highly suitable for studying the energy landscape of defects in semiconducting thin films \cite{Lonsky2016, Lonsky2015} and semiconductor heterostructures \cite{Mueller2006, Mueller2006a, Muller2015}.      

\section{Experimental Details}
\subsection{Sample Growth} \label{growth}
Electronic transport measurements have been performed on a total of seven Ga$_{1-x}$Mn$_{x}$As thin film samples and one Ga$_{1-x}$Mn$_{x}$P sample, see Table \ref{tab:table1} for an overview. For all films, the corresponding Curie temperatures were determined by magnetization measurements \cite{Zhou2016, Hamida2014, Yuan2017a, Zhou2015}.
In general, a crucial parameter is the Mn content $x$, but other factors as thermal annealing or induced disorder by ion irradiation also have a strong influence on the Curie temperature $T_\mathrm{C}$ and the crystalline defect characteristics \cite{Hayashi2001, Potashnik2001, Esch1997, Sinnecker2010}.
The samples in this study were prepared in two different ways. Metallic Ga$_{1-x}$Mn$_{x}$As samples with $x=4$ \% were grown by low-temperature molecular beam epitaxy (LT-MBE) on semi-insulating GaAs(001) substrates in a Mod Gen II MBE system with the lowest possible As\textsubscript{2}-partial pressure of about $2\times 10^{-6}\,$mbar at PTB in Braunschweig \cite{Hamida2014}. After the growth of a $100\,$nm high-temperature (HT) GaAs buffer layer at $T_\mathrm{g}=560\,^\circ$C, the temperature was decreased to $270\,^\circ$C for the subsequent LT Ga$_{1-x}$Mn$_{x}$As growth. Post-growth annealing at $200\,^\circ$C (18 hours in ambient atmosphere) was performed for one of the samples in order to enhance $T_\mathrm{C}$. The total Mn concentration $x$ was calculated from the molecular flux ratio of Mn and Ga measured in the MBE at the position of the wafer and compared with reflection high-energy electron diffraction (RHEED) and energy-dispersive X-ray spectroscopy (EDX) measurements.   
Moreover, three Ga$_{1-x}$Mn$_{x}$As samples with a nominal Mn concentration of $x=6$ \% were grown by LT-MBE on semi-insulating GaAs(001) using a Veeco Mod III MBE system in Nottingham \cite{Zhou2016, Wang2008}. In this case, thermal annealing was performed at $190\,^\circ$C for 48 hours in ambient atmosphere and the Mn content was determined from the Mn/Ga flux ratio. Two of the films were irradiated with different doses of He ions after growth. This particular method allows to control the hole concentration and thus the electronic as well as the magnetic properties without changing the Mn content of a sample. The He-ion energy was chosen as $4\,$keV. The fluences were $2.5\times 10^{13}\,$cm$^{-2}$ and $3.5\times 10^{13}\,$cm$^{-2}$ for the two irradiated samples. A better measure for the effect of irradiation on material properties than the fluence is the so-called displacement per atom (DPA), i.e.\ the number of times that an atom in the target is displaced during irradiation. This allows for a comparison with data reported in the literature, in which other ion species and energies are used. For the two irradiated samples, the DPA was $1.6\times 10^{-3}$ and $2.24\times 10^{-3}$, respectively \cite{Zhou2016}. During ion irradiation, the films were tilted by $7^{\circ}$ to avoid channeling. The irradiation parameters result in defects distributed roughly uniformly in the whole Ga$_{1-x}$Mn$_{x}$As layer as confirmed by simulations using the SRIM (Stopping and Range of Ions in Matter) code \cite{Ziegler1985}. No measurable increase of Mn interstitials was observed by Rutherford backscattering spectroscopy (RBS) \cite{Zhou2016}. Previous studies show that also the sheet concentration of substitutional Mn atoms remains constant \cite{Winkler2011}, which is why we conclude that the main effect of He-ion irradiation is to introduce deep traps and thereby compensate the holes. It is well established that these defects reside in the As sublattice and most of them are primary defects related to vacancies and interstitials \cite{Sinnecker2010, Pons1985}. In our case, atomic force microscopy (AFM) and RBS measurements do not show any indications of irradiation induced surface reconstruction. Further details can be found in Ref.\ \cite{Zhou2016}.    
Finally, two Ga$_{1-x}$Mn$_{x}$As samples with low Mn contents of $1.8$ \% and $1.2$ \% as well as a Ga$_{1-x}$Mn$_{x}$P sample with $x=3.5$ \% were fabricated by ion implantation combined with pulsed laser melting in Dresden \cite{Yuan2017a, Zhou2015}. Ion implantation is a common materials engineering technique for introducing foreign ions into a host material. In this case, Mn ions are implanted into GaAs or GaP wafers. The subsequent laser pulse drives a rapid liquid-phase epitaxial growth. The implantation energy was set to $100\,$keV for GaAs \cite{Yuan2017a} and $50\,$keV for GaP \cite{Yuan2016, Zhou2015}. The wafer normal was tilted by $7^{\circ}$ with respect to the ion beam to avoid a channeling effect. A coherent XeCl laser (with $308\,$nm wavelength and $28\,$ns duration) was employed to recrystallize the samples, and the energy densities were optimized to achieve high crystalline quality and the highest Curie temperature. The optimal laser energy density is $0.30\,$J/cm$^2$ for Ga$_{1-x}$Mn$_{x}$As and $0.45\,$J/cm$^2$ for Ga$_{1-x}$Mn$_{x}$P. The Mn concentration was determined by secondary ion mass spectroscopy. In contrast to films grown by LT-MBE, neither Mn interstitials nor As antisites are observed in samples prepared by ion implantation combined with pulsed laser melting \cite{Yuan2017a, Cho2007}. Transmission electron microscopy (TEM) studies prove the high crystalline quality of the films and exclude the presence of any extended lattice defects, amorphous inclusions and precipitates of other crystalline phases \cite{Yuan2017a}.     
For some selected films, an array of $50\times 50\, \mu$m$^2$ Hall bars was defined by photolithography followed by wet chemical etching. The quality of Hall effect measurements thereby improves significantly due to a well-defined contact geometry and, as explained in more detail below, the resistance noise magnitude as the desired measurement signal increases due to smaller sample volumes according to Hooge's law \cite{Hooge1969, Hooge1976}. For all samples, electrical contacts were made by soldering In/Sn on top of the films.
Charge carrier concentrations obtained from Hall effect measurements are given in Table \ref{tab:table1}. Due to the use of two different fabrication techniques and since the LT-MBE samples from Braunschweig and Nottingham were grown at different substrate temperatures and As-fluxes, care has to be taken in comparing the values for the hole concentration $p$ of films of different origin. Apart from that, as expected, the hole density increases after thermal annealing for the $x=4$ \% samples and decreases with increasing ion irradiation dose for the $x=6$ \% samples, cf.\ Section \ref{ssecmetallic} for more details. Furthermore, there is also a clear correlation between $T_\mathrm{C}$ and $p$. Finally, we point out that extensive studies on all present samples, including magnetization measurements and standard thin film characterization techniques, have been published elsewhere, cf.\ Ref.\ \cite{Hamida2014, Zhou2016, Yuan2017a, Zhou2015}.      
\begin{table*}
\caption{\label{tab:table1}Overview of the investigated thin film samples and related parameters, including: information about the manganese content $x$, whether samples were grown by low-temperature molecular beam epitaxy (LT-MBE) or ion implantation combined with pulsed laser melting (II+PLM), the institute where samples have been fabricated, which kind of post-treatment was given, the values of the film thickness, the Curie temperature $T_\mathrm{C}$ as determined by magnetization measurements, and the hole density $p(T=300\,\mathrm{K})$ obtained from Hall effect measurements.}
\begin{ruledtabular}
\begin{tabular}{ccccccc}
Mn content&Fabrication&Source&Remarks&Thickness&$T_\mathrm{C}$ & $p(T=300\,\mathrm{K})\ [1/\mathrm{cm}^3]$\\
\hline
Ga$_{1-x}$Mn$_{x}$As\\
%7.0 \% & LT-MBE & Braunschweig & as-grown & $25\,$nm & $60\,$K & $3.7\times 10^{19}$\\
%7.0 \% & LT-MBE & Braunschweig & annealed at 200$^\circ$C (18 hours)& $25\,$nm & $122\,$K & $1.3\times 10^{20}$\\
4.0 \% & LT-MBE & PTB & as-grown, micro-structured& $25\,$nm & $70\,$K & $8.0\times 10^{19}$\\
4.0 \% & LT-MBE & PTB & annealed ($200\,^\circ$C, $18\,$h), micro-structured& $25\,$nm & $110\,$K & $1.2\times 10^{20}$\\
6.0 \% & LT-MBE & Nottingham & annealed ($190\,^\circ$C, $48\,$h)& $25\,$nm & $125\,$K & $9.6\times 10^{20}$\\
6.0 \% & LT-MBE & Nottingham & annealed, He-ion irradiated (low dose)& $25\,$nm & $75\,$K & $8.0\times 10^{20}$\\
6.0 \% & LT-MBE & Nottingham & annealed, He-ion irradiated (high dose)& $25\,$nm & $50\,$K & $5.6\times 10^{20}$\\
1.8 \% & II+PLM & HZDR & as-grown, micro-structured& $60\,$nm & $60\,$K & $2.8\times 10^{20}$\\
1.2 \% & II+PLM & HZDR & as-grown, micro-structured& $60\,$nm & $31\,$K & $1.0\times 10^{20}$\\
%0.9 \% & II+PLM & Dresden & as-grown& $60\,$nm & $17\,$K & $2.9\times 10^{20}$\\
\hline
Ga$_{1-x}$Mn$_{x}$P\\
3.5 \% & II+PLM & HZDR & as-grown& $34\,$nm & $45\,$K & $3.1\times 10^{20}$\\
\end{tabular}
\end{ruledtabular}
\end{table*}

\subsection{Measurements}
\begin{figure}%[b]
\centering
\includegraphics[width=8.5 cm]{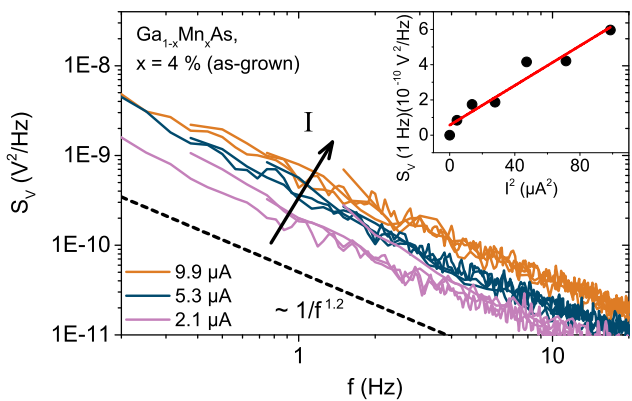}
\caption{Exemplary current-dependent noise spectra $S_{V}(f)$ acquired for the as-grown Ga$_{1-x}$Mn$_{x}$As sample with $x=4$ \%. Inset: A quadratic dependence (red line: linear fit to the data) between $S_{V}$ and $I$ verifying Hooge's empirical law.} 
\label{PTB_CURRENT_DEP}%
\end{figure}

Electronic transport measurements have been performed using both AC and DC techniques. Experiments were carried out in a continuous-flow cryostat with variable temperature insert. Magnetic fields were applied perpendicular to the film plane.
For some of the lithographically patterned Ga$_{1-x}$Mn$_{x}$As samples, low-frequency noise spectroscopy was conducted in a five-terminal AC setup, where the sample is placed in a bridge-circuit in order to suppress the constant DC voltage offset and to minimize external perturbations \cite{Scofield1987}. Other samples were measured in a four-terminal AC or DC setup. As a few samples showed a frequency-dependent resistivity at low temperatures, the excitation frequency was reduced from $227\,$Hz down to $17\,$Hz, or a DC noise measurement setup was utilized, which was verified to yield the same results as AC noise measurements.        
The fluctuating voltage signal is preamplified and processed by a spectrum analyzer yielding the voltage noise power spectral density (PSD) $S_{V}(\omega)$ defined by:
\begin{equation}
S_{V}(\omega)=2\lim\limits_{T \to \infty}\frac{1}{T}\left| \int_{-T/2}^{T/2} {\rm d}t \, e^{-i\omega t} \, \delta V(t) \right|^2,
\end{equation}        
where $\delta V(t)$ represents the fluctuating voltage drop across the sample and $\omega = 2\pi f$ the angular frequency. 
Care was taken that all spurious sources of noise were minimized or eliminated. 
As required by Hooge's empirical law \cite{Hooge1969, Hooge1976},
\begin{equation}
S_{V}(f)=\frac{\gamma_{H}\cdot V^2}{n \Omega f^{\alpha}},
\end{equation}   
the magnitude of the voltage noise scales as $S_V \propto V^2 \propto I^2$, where $I$ represents the current flowing through the sample. 
Here, $n$ is the charge carrier density and $\Omega$ the 'noisy' sample volume, i.e.\ $n \Omega = N_c$ gives the total number of  charge carriers in the material causing the observed $1/f$ noise. $\alpha$ describes the frequency exponent which is commonly in the range $0.8\leq \alpha \leq 1.4$ for $1/f$-type fluctuations.  
The Hooge parameter $\gamma_H$ is widely used to compare the noise level of different systems and covers a range of $\gamma_H=10^{-6}$--$10^{7}$ for different materials \cite{Raquet2001}. For bulk semiconductors, $\gamma_H$ usually is of order $10^{-2}$--$10^{-3}$. Moreover, it is useful to normalize the magnitude of the voltage fluctuations with respect to the applied current, $S_{R}=S_{V}/I^2$, and to the resistance of the sample, resulting in $S_{R}/R^2$.
Exemplary noise spectra for a Ga$_{1-x}$Mn$_{x}$As sample are presented in Fig.$\,$\ref{PTB_CURRENT_DEP} for three different currents in a log-log plot. The dashed line represents a $1/f^{\alpha}$ function with a slope of $\alpha =1.2$. The inset clarifies the quadratic scaling of $S_{V}$ with the current $I$ as predicted by Hooge's law. 
Further details about the fluctuation spectroscopy technique can be found in \cite{Raquet2001,Mueller2011}.       

\section{Results and Discussion}

\subsection{Ga\textsubscript{1-x}Mn\textsubscript{x}As films with high Mn content} \label{ssecmetallic}

\begin{figure}%[b]
\centering
\includegraphics[width=8.5 cm]{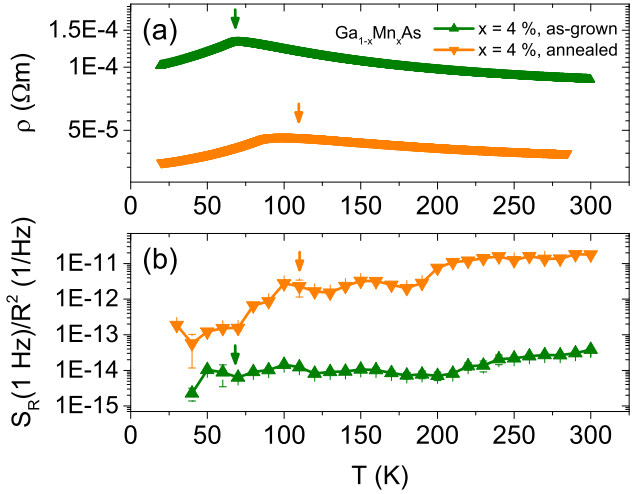}
\caption{(a) Temperature-dependent resistivity showing characteristic maxima around $T_\mathrm{C}$ (marked by arrows) and (b) normalized resistance noise magnitude of as-grown and annealed Ga$_{1-x}$Mn$_{x}$As samples with 4 \% Mn content. No features in the noise power are visible around $T_\mathrm{C}$.} 
\label{PTB_PSD}%
\end{figure}

At first, we will discuss measurements on metallic films with Mn contents $x>2$ \%. Resistivity curves of as-grown and annealed Ga$_{1-x}$Mn$_{x}$As samples with $x=4$ \% are depicted in Fig.$\,$\ref{PTB_PSD}(a). As expected, typical maxima \cite{Matsukura1998} are observed in the vicinity of the samples' respective Curie temperature $T_\mathrm{C}$. After post-growth annealing, the resistivity significantly decreases while $T_\mathrm{C}$ increases compared to as-grown films. This effect can be explained by the out-diffusion of Mn interstitials to the film surface \cite{Edmonds2004}. Commonly, these interstitials act as donors and compensate the hole-mediated ferromagnetism. 
Fig.$\,$\ref{PTB_PSD}(b) shows the corresponding normalized magnitude of the resistance fluctuations $S_{R}/R^2$ at $1\,$Hz as a function of temperature in a semi-logarithmic representation. Remarkably, the PSD varies over several orders of magnitude for the two films and shows, in contrast to the resistivity, no significant features around $T_\mathrm{C}$. More specifically, there are no major changes in the noise behavior throughout the entire temperature range. In addition, a constant external out-of-plane magnetic field up to $7\,$T does not lead to any changes in the normalized resistance noise (not shown, cf.\ Ref.\ \cite{Lonsky2017}). Possible contributions from mixed phases could be overshadowed by fluctuations related to thermally-activated impurity switching processes. However, more likely, due to the high concentration of Mn substitutional atoms, charge carriers are delocalized and the formation of magnetic polarons is not to be expected for these metallic samples. The observed $1/f$-type noise is likely to be dominated by switching processes related to crystalline defects, which can also be seen in a strong variation of the Hooge parameter $\gamma_H$ for the two different samples. At room temperature, we obtain $\gamma_H=1\times 10^{-2}$ for the as-grown and $\gamma_H=3\times 10^0$ for the annealed film. Apparently, thermal annealing has a strong influence on the noise magnitude, leading to the presumption that slow fluctuation processes related to Mn interstitials might play an important role. Since thermal annealing reduces the density of Mn interstitials, one might expect a lower PSD for annealed samples, but the opposite is the case. The same behavior is observed for two $x=7$ \% samples with a thickness of $d=25\,$nm (not shown). Strikingly, for two further $x=4$ \% films with a higher thickness of $d=100\,$nm and similar values of $T_{\mathrm{C}}$, a slight decrease of the noise magnitude is observed after thermal annealing, indicating that the rearrangement of Mn interstitials due to diffusion processes towards the surface \cite{Edmonds2004}, which become passivated due to oxidation or by binding with surplus As atoms, and concomitant surface effects may play an important role for the changes in the PSD after thermal annealing.
   
\begin{figure}%[b]
\centering
\includegraphics[width=8.5 cm]{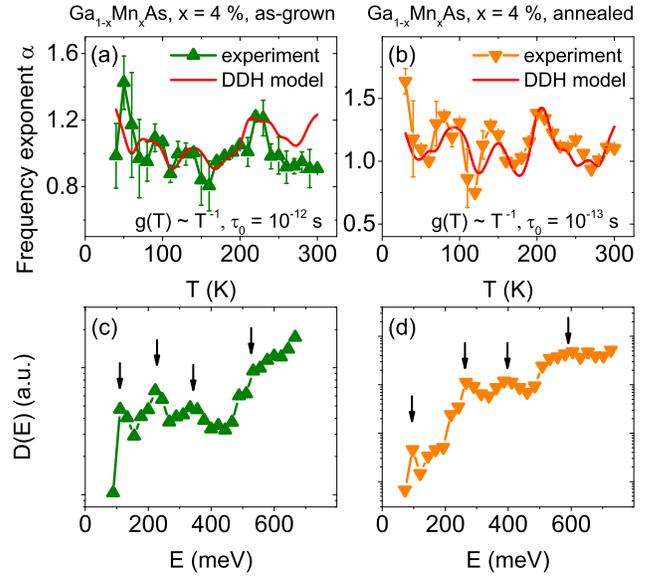}
\caption{(a) and (b) Application of the phenomenological DDH model (red curves) to the resistance noise data of two metallic Ga$_{1-x}$Mn$_{x}$As samples with  $x=4$ \%, both showing a reasonable agreement between calculated and measured values for the frequency exponent $\alpha(T)$. The function $g(T)$ and the attempt time $\tau_0$ are indicated in each case. (c) and (d) Calculated distribution of activation energies $D(E)$ for both samples, showing a similar trend. Local maxima are indicated by black arrows.}
\label{PTB_DDH}%
\end{figure}

In order to deduce the characteristic energies of the switching processes contributing to the $1/f$-type noise, we apply the phenomenological model of Dutta, Dimon and Horn (DDH) \cite{Dutta1979}. In this model, a certain distribution of activation energies $D(E,T)$ determines the temperature dependence of both the noise magnitude and the frequency exponent. An essential requirement for the applicability of this model is to check whether $\alpha(T)$ calculated after
\begin{equation} \label{eq:DDH}
\alpha(T)=1-\frac{1}{\ln(2\pi f\tau _0)} \left[\frac{\partial \ln(\frac{S_R}{R^2}(f,T))}{\partial \ln(T)}-g^{\prime}(T)-1\right]
\end{equation}   
%&
agrees with the measured values. Here, $\tau _0$ represents an attempt time, usually between $10^{-14}$ and $10^{-11}\,$s, corresponding to typical inverse phonon frequencies. Moreover, it is $g^{\prime}(T)=\frac{\partial \ln(g(T))}{\partial \ln(T)} \equiv b$, where $g(T)=a \cdot T^{b}$ accounts for an explicit temperature dependence of the distribution of activation energies $D(E,T)$, which can be caused by a change of the number of thermally activated switching events (excitation of defect states) or of the coupling of fluctuations to the resistivity with temperature. As can be seen in Fig.$\,$\ref{PTB_DDH}(a) and (b), we find a good qualitative or even quantitative agreement between model and experiment for both $x=4$ \% samples. This allows for calculating the distribution of activation energies,
\begin{equation}
D(E) \propto \frac{2\pi f}{k_B T}\frac{1}{g(T)}\frac{S_R}{R^2}(f,T), 
\end{equation}
of thermally activated fluctuators, which is shown in Fig.$\,$\ref{PTB_DDH}(c) and (d). Here, it is $E=-k_B T \ln(2\pi f \tau_{0})$. Both samples show a similar behavior, namely an increase of $D(E)$ towards higher activation energies, which we interpret as a superposition of several thermally activated processes with different energies, that can be attributed to various kinds of defects. In both cases, four local maxima in $D(E)$ are observed and marked by black arrows in Fig.$\,$\ref{PTB_DDH}(c) and (d), whereby the values of the corresponding activation energies are remarkably similar for the two samples. It is plausible to assume that, due to the low growth temperatures utilized during LT-MBE, which are required in order to prevent phase separation within the material, a great variety of defects, such as Mn interstitials and As antisites, contribute to the distribution of activation energies. Although the energies of the local maxima are comparable to typical impurity binding energies in GaAs, care has to be taken when assigning the energies to specific defect states due to band gap renormalization in heavily doped semiconductors, which is accompanied by a shift of the respective binding energies.       

\begin{figure}%[b]
\centering 	
\includegraphics[width=8.5 cm]{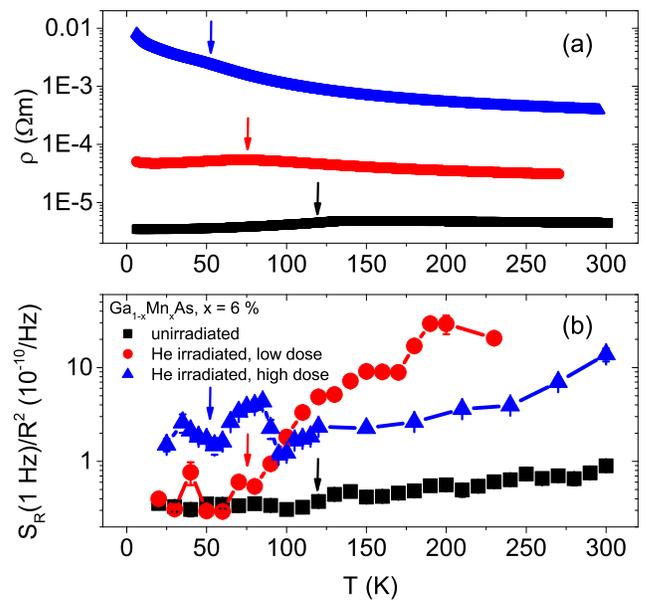}
\caption{(a) Temperature-dependent resistivity data of three Ga$_{1-x}$Mn$_{x}$As ($x=6$ \%) samples with different He-ion irradiation doses. (b) Normalized noise PSD as a function of temperature for the three films in a logarithmic representation. No magnetic field dependence or significant features around $T_\mathrm{C}$ were observed. Arrows indicate the corresponding Curie temperatures.} 
\label{GAMNAS_RES}%
\end{figure}

Since no signatures of electronic phase separation can be identified for conventional metallic Ga$_{1-x}$Mn$_{x}$As samples with high Mn content ($x=4$ \%), we next focus on samples irradiated by He ions, whereby disorder in the films is enhanced by the introduction of deep traps, i.e.\ the Fermi level is shifted by means of carrier compensation in order to change the conductivity from metallic to insulating.     
Fig.$\,$\ref{GAMNAS_RES}(a) shows resistivity data of three Ga$_{1-x}$Mn$_{x}$As samples with $x=6$ \%. Point defects were introduced by the irradiation with an energetic He-ion beam using different doses as described in Section \ref{growth} (fluences: $2.5\times 10^{13}\,$cm$^{-2}$ and $3.5\times 10^{13}\,$cm$^{-2}$, hereinafter referred to as "low dose" and "high dose", respectively). The corresponding Curie temperatures determined from magnetization measurements are marked by arrows. While the unirradiated sample shows the lowest resistivity and metallic behavior with a relatively high $T_\mathrm{C}$ of $125\,$K, the resistivity increases strongly as a function of He-ion irradiation dose. At the same time, $T_\mathrm{C}$ decreases down to $75\,$K for the sample irradiated with a low He-ion dose and even further to $50\,$K for a high irradiation dose. The typical maximum in resistivity \cite{Edmonds2002} becomes less pronounced for more insulating samples. These major changes in resistivity can be explained by an increase of the displacement per atom (DPA) for higher irradiation doses, which results in a decrease of the hole concentration. This is confirmed by Hall effect measurements at room temperature for the three samples (cf.\ Table \ref{tab:table1}).
The normalized temperature-dependent resistance noise PSD $S_{R}/R^2$ at $1\,$Hz for the $x=6$ \% samples is shown in Fig.$\,$\ref{GAMNAS_RES}(b) in a semi-logarithmic plot. The unirradiated sample has the lowest noise level over the entire temperature range. At temperatures below $100\,$K, the PSD is nearly independent of temperature, followed by a slight increase above $100\,$K towards room temperature. Likewise, below $80\,$K, the film irradiated with a low He-ion dose shows the same constant noise magnitude, while there is a much stronger increase of nearly two orders of magnitude towards higher temperatures. The sample irradiated with the high dose shows the highest noise level of all three films below $100\,$K, also followed by a characteristic increase for $T$ approaching room temperature. The great variation in the PSD for the different samples can be explained by the introduction of deep traps into the As sublattice. An exchange of charge carriers between such traps and the rest of the conducting material, resulting in fluctuations of the hole concentration, can cause the observed $1/f$-type noise. As shown in previous studies \cite{Zhou2016}, the concentration of Mn interstitials should not change as a function of irradiation dose, which implies that these interstitials are not the cause for the variation in the PSD as a function of He-ion irradiation dose, although they might still contribute to the $1/f$ noise. In addition, the noise data shown in Fig.$\,$\ref{GAMNAS_RES}(b) suggest a crossover between two temperature regimes: A temperature-independent region below about $100\,$K and a characteristic increase of the noise magnitude, where the number of activated defects increases towards higher temperatures. The Hooge parameter $\gamma_{H}$ at room temperature for these samples is of order $\gamma_{H}=10^{3}$--$10^{5}$ and therefore several orders of magnitude larger than for 'clean' semiconductors. Furthermore, $\gamma_{H}$ is also several orders of magnitude higher than for the $x=4$ \% samples, presumably due to the higher Mn content, a higher concentration of traps in the case of the irradiated samples and different growth parameters (substrate temperature and annealing procedure). \newline 
Apart from that, no features around $T_\mathrm{C}$ were observed except a slight increase below $T_\mathrm{C}$ for the highest He-ion dose, which might be a hint for weak electronic phase separation or the increasing localization of charge carriers, and a pronounced peak at $80\,$K. As shown in Ref.\ \cite{Lonsky2017}, the application of the phenomenological model by Dutta, Dimon and Horn \cite{Dutta1979} allows to assign the enhanced noise magnitude at temperatures between 60 and 100\,K to a distinct peak in the distribution of activation energies $D(E)$ at energies of about $180\,$meV. This energy can very likely be attributed to traps introduced by the strong He-ion irradiation. No changes in the normalized noise power were found in external magnetic fields up to $7\,$T (cf.\ Ref.\ \cite{Lonsky2017}). Except the above-mentioned weak increase of the PSD below $T_\mathrm{C}$ for the sample irradiated with the high dose, there are no indications for an electronic phase separation. As suggested by the large values of $\gamma_{H}$, possible contributions attributed to a percolative magnetic phase transition may be overshadowed by disorder effects or impurity switching processes. A different approach in order to find indications for a possible electronic phase separation is to tune the Mn content of Ga$_{1-x}$Mn$_{x}$As, which will be discussed in the following section.

\subsection{Ga\textsubscript{1-x}Mn\textsubscript{x}As films with low Mn content and Ga\textsubscript{1-x}Mn\textsubscript{x}P}

\begin{figure}%[b]
\centering
\includegraphics[width=8.5 cm]{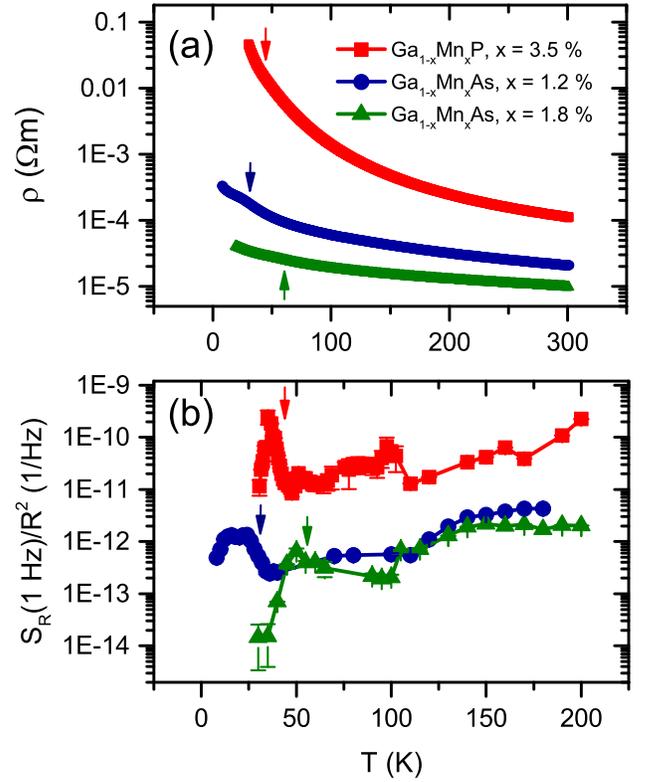}
\caption{(a) Resistivity curves of two Ga$_{1-x}$Mn$_{x}$As samples with low Mn content ($x=1.8$ \% and $x=1.2$ \%) and a Ga$_{1-x}$Mn$_{x}$P reference sample ($x=3.5$ \%). (b) Temperature-dependent noise magnitude of the three samples, all showing an enhanced $1/f$ noise just below $T_\mathrm{C}$. Curie temperatures are marked by arrows.} 
\label{HZDR_PSD_IN}%
\end{figure}

\begin{figure}%[b]
\centering
\includegraphics[width=8.5 cm]{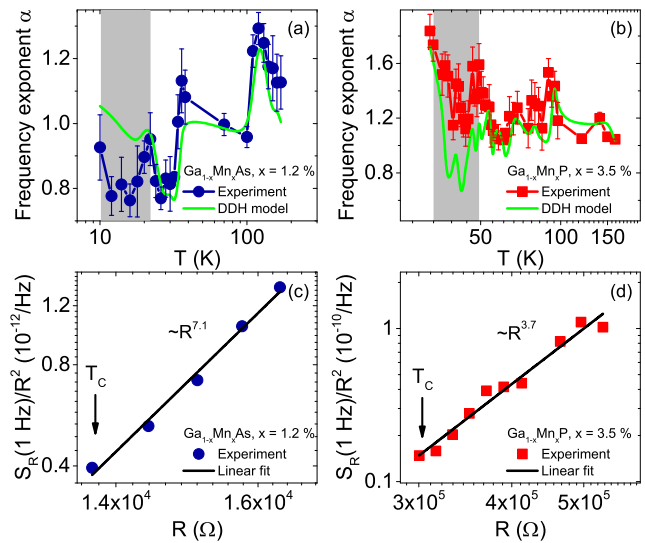}
\caption{(a) and (b) Comparison between experimentally determined and calculated (DDH model) frequency exponents for the Ga$_{1-x}$Mn$_{x}$As sample with $x=1.2$ \% and the Ga$_{1-x}$Mn$_{x}$P film. Strong deviations occur in the gray shaded areas around the percolative transition. (c) and (d) Log-log plot of PSD versus resistance for Ga$_{1-x}$Mn$_{x}$As ($x=1.2$ \%)  between $24$ and $32\,$K and for Ga$_{1-x}$Mn$_{x}$P between $35$ and $45\,$K. The solid black lines correspond to linear fits yielding $S_{R}/R^2\propto R^{w}$ with $w=7.1\pm 0.3$ for the weakly doped Ga$_{1-x}$Mn$_{x}$As film and  $w=3.7\pm 0.3$ for the Ga$_{1-x}$Mn$_{x}$P sample.} 
\label{HZDR_GAMNAP_ANALYS}%
\end{figure}

In this section, we focus on weakly doped Ga$_{1-x}$Mn$_{x}$As samples with localized charge carriers and compare the results with the metallic films and another insulating Ga$_{1-x}$Mn$_{x}$P reference sample. Fig.\ \ref{HZDR_PSD_IN}(a) depicts the resistivity of two as-grown Ga$_{1-x}$Mn$_{x}$As samples with $x=1.8$ \% and $x=1.2$ \% and a Ga$_{1-x}$Mn$_{x}$P film with $x=3.5$ \%. In contrast to the metallic samples, no or only weak features around $T_\mathrm{C}$ are visible in the resistivity. However, for all three films, the $1/f$ noise magnitude is significantly enhanced just below the respective Curie temperature, as can be seen in Fig.$\,$\ref{HZDR_PSD_IN}(b), where the normalized resistance noise at $1\,$Hz is plotted versus temperature. For the $x=1.8$ \% film, only a weak increase occurs around $T_\mathrm{C}$, followed by a sharp decrease towards lower temperatures. Although this compound is still to be considered as metallic, in the phase diagram it is located very close to the metal-insulator transition (MIT) \cite{Yuan2017a}, which is why weak signatures of an electronic phase separation are conceivable. The Ga$_{1-x}$Mn$_{x}$As film with $x=1.2$ \% exhibits nearly the identical noise magnitude between $50$ and $170\,$K, which is easily comprehensible since the same fabrication technique has been employed, the Mn content is very similar and hence the defect landscape contributing to the resistance noise is comparable. The calculated Hooge parameters at room temperature for both samples are also comparable, namely  $\gamma_{H}=2\times 10^{0}$ and $\gamma_{H}=5\times 10^{1}$ for the $x=1.2$ \% and the $x=1.8$ \% films, respectively. However, the peak in the vicinity of $T_\mathrm{C}$ is much more pronounced for the $x=1.2$ \% sample, because this compound is situated right on the edge of the metal-insulator transition, cf.\ studies of electrical and magnetic properties on the present samples in Ref.\ \cite{Yuan2017a}. In detail, this sample still exhibits a global ferromagnetic behavior below $T_\mathrm{C}$, but electronic transport measurements indicate its insulating character. Within the framework of this study, no noise measurements on Ga$_{1-x}$Mn$_{x}$As samples situated on the insulating side of the MIT could be performed, because the maximum possible current $I$ was not sufficient to measure $1/f$-type spectra reliably. Instead, the investigated Ga$_{1-x}$Mn$_{x}$P sample is suggested to provide a reference example for the signatures of an electronic phase separation in fluctuation spectroscopy measurements of a diluted magnetic semiconductor with localized charge carriers. For this sample, the Hooge parameter at room temperature amounts to $\gamma_{H}=1\times 10^{4}$. As can be seen in Fig.\ \ref{HZDR_PSD_IN}(b), the Ga$_{1-x}$Mn$_{x}$P film exhibits a pronounced peak just below $T_\mathrm{C}$, where the noise level increases by more than one order of magnitude in a small temperature interval.        
In analogy to previous studies on the semimetallic ferromagnet EuB\textsubscript{6} \cite{Das2012}, the diverging behavior of the resistance noise PSD for the Ga$_{1-x}$Mn$_{x}$P sample can be described by a Lorentz function with a peak at $35.5\,$K and a width $\Delta T=2.5\,$K. In the case of EuB\textsubscript{6}, Das et al.\ attribute this sharp peak to a magnetic polaron percolation. As suggested by Kaminski and Das Sarma, such a behavior is also to be expected for Ga$_{1-x}$Mn$_{x}$As or Ga$_{1-x}$Mn$_{x}$P samples with strongly localized charge carriers \cite{Kaminski2002, Kaminski2003}. Due to the high defect concentration in DMS, it is assumed that the charge carrier concentration is highly inhomogeneous and as ferromagnetism is mediated by charge carriers, upon decreasing the temperature, the ferromagnetic transition will first occur locally within the regions with higher carrier concentration. Upon lowering the temperature, these finite-size clusters will grow and merge until the entire sample becomes ferromagnetic via a percolation transition. 
However, in contrast to semimetallic EuB\textsubscript{6}, where the ferromagnetic transition is accompanied by a drastic reduction of $\rho(T)$ and a colossal magnetoresistance effect, the temperature dependence of the resistivity of the investigated insulating DMS samples is monotonic. For samples located in the vicinity of the metal-insulator transition, only a small kink is observable in $\rho(T)$ around $T_\mathrm{C}$. In general, a ferromagnetic percolation transition is accompanied by an increase of the electrical conductivity, but, at the same time, this may be compensated by the increase of the resistivity with decreasing temperature due to the semiconducting nature of the material \cite{Kaminski2003}. In the case of strongly localized charge carriers, the decrease in their hopping rate upon decreasing temperature overcomes the decrease in the hopping activation energy due to the ferromagnetic transition. Strikingly, although no features can be observed in the resistivity at $T_\mathrm{C}$, resistance noise, which is very sensitive to the microscopic current distribution in the sample, shows a strong peak for the Ga$_{1-x}$Mn$_{x}$P sample. Due to the less localized nature of holes in the present weakly doped Ga$_{1-x}$Mn$_{x}$As samples, the possible percolation transition and thus the enhancement of the PSD are less pronounced. In addition, in the vicinity of the maximum in $S_{R}/R^2$, we find a strong deviation between the calculated frequency exponents from Eq.\ (\ref{eq:DDH}) and the experimentally determined values, which is shown in Fig.\ \ref{HZDR_GAMNAP_ANALYS}(a) and (b) for the Ga$_{1-x}$Mn$_{x}$As ($x=1.2$ \%) and the Ga$_{1-x}$Mn$_{x}$P samples, respectively. This non-applicability of the DDH model (marked by gray shaded areas) is another indication for a percolative transition \cite{Kogan1996}, since the assumptions of this phenomenological approach are not compatible with the nonlinear electronic transport behavior around the percolation threshold. The deviations between calculated and experimentally determined frequency exponents and the divergence in the PSD in the vicinity of the percolation threshold $p_c$ can be understood within the frame of a random resistor network (RRN) model \cite{Kogan1996}. The reduced number of effective current paths results in the suppression of cancellation of uncorrelated resistance fluctuations along different paths, which are abundant far away from $p_c$ \cite{Das2012}. Around $p_c$ the current density is strongly inhomogeneous and the most significant contribution to the resistance noise comes from so-called bottlenecks which connect large parts of the infinite cluster. Here, the current density is higher than in other parts of the network.             
Rammal et al.\ have shown that near the percolation threshold $p_c$, the PSD diverges as $S_{R}/R^2\propto(p-p_c)^{-\kappa}$, while the resistance $R$ behaves as $R\propto (p-p_c)^{-t}$ \cite{Rammal1985}. Here, $\kappa$ and $t$ are critical percolation exponents derived from a RRN model, and $p$ is the fraction of unbroken bonds of a RRN. Due to the non-accessibility of these exponents in an experiment, it is common to link the PSD and the resistance via $S_{R}/R^2 \propto R^{w}$ with $w = \kappa /t$ \cite{Yagil1992}. The corresponding analysis for the two relevant thin films is shown in Fig.$\,$\ref{HZDR_GAMNAP_ANALYS}(c) and (d), yielding a critical exponent $w=3.7\pm 0.3$ for Ga$_{1-x}$Mn$_{x}$P and $w=7.1\pm 0.3$ for Ga$_{1-x}$Mn$_{x}$As ($x=1.2$ \%). While for Ga$_{1-x}$Mn$_{x}$P this is in fair agreement with typical values for the exponent $w$, e.g.\ $w=2.9\pm 0.5$ for perovskite manganites \cite{Podzorov2000}, the value for the Ga$_{1-x}$Mn$_{x}$As ($x=1.2$ \%) sample is exceptionally high. We note that no clear systematic changes of shape, position and height of the peak in the temperature-dependent PSD as a function of the applied out-of-plane magnetic field $B$ can be observed. It is assumed that possible changes as a function of the external field are too weak in order to be resolved.             
We emphasize that the discussed picture of percolating bound magnetic polarons is expected to be only valid within the impurity-band model and is not compatible with the p-d Zener model, which assumes ferromagnetism being mediated by a Fermi sea of itinerant holes. The impurity-band model suggests that even for strong Mn doping the Fermi energy is located within the separate impurity band and only the degree of localization of the charge carriers will change. 
Our results support the view of holes being trapped in localized impurity band states for weakly doped Ga$_{1-x}$Mn$_{x}$As as well as for Ga$_{1-x}$Mn$_{x}$P, whereas for metallic Ga$_{1-x}$Mn$_{x}$As samples with higher $x$, where no signatures of a percolation mechanism were observed in the electronic noise, the widely-held view of delocalized holes within the valence band mediating ferromagnetism is more appropriate. We note that, for instance, DC transport and optical studies \cite{Jungwirth2007} as well as first-principle calculations \cite{Bae2016} corroborate the applicability of the two different models on Ga$_{1-x}$Mn$_{x}$As for the respective Mn concentration ranges. Moreover, because of a strong variation of the Mn energy level among different III-Mn-V combinations, it is unlikely that all materials can be treated within a single model \cite{Khalid2014}. As a consequence of the higher degree of hole localization in Ga$_{1-x}$Mn$_{x}$P, the peak in $S_{R}/R^2$ is more pronounced as compared to the insulating Ga$_{1-x}$Mn$_{x}$As sample.            
An alternative interpretation of our findings for the Ga$_{1-x}$Mn$_{x}$As with small $x$ valid within the framework of the p-d Zener model could be an electronic phase separation in the vicinity of the metal-insulator transition  \cite{Ohno2008, Dietl2006}, cf.\ Section \ref{intro}. In this case, the nano-scale phase separation results in the coexistence of ferromagnetic bubbles (metallic, hole-rich regime) and a paramagnetic matrix (insulating, hole-poor regime). However, this kind of electronic phase separation should persist in a broad temperature range below $T_\mathrm{C}$ for insulating Ga$_{1-x}$Mn$_{x}$As samples with low Mn doping, which is not expected to result in such a pronounced and sharp peak in the temperature-dependent noise power spectral density as it was observed in this work. It should also be noted that all investigated samples show global ferromagnetism below $T_\mathrm{C}$, i.e.\ there are no mixed phases consisting of ferromagnetic clusters and superparamagnetic grains. Yuan et al.\ have shown that these mixed phases exist for Ga$_{1-x}$Mn$_{x}$As samples with $x\leq 0.9$ \%, but not for $x\geq 1.2$ \% \cite{Yuan2017a}. It is desirable to study the resistance noise behavior of such mixed phases in future.      

\section{Summary and Conclusion}
In this work, we investigated the resistance noise on a series of Ga$_{1-x}$Mn$_{x}$As films with different manganese and defect concentrations and an insulating Ga$_{1-x}$Mn$_{x}$P reference sample, all of which exhibit global ferromagnetism below $T_\mathrm{C}$. By applying the phenomenological model by Dutta, Dimon and Horn, we calculated the distribution of activation energies $D(E)$ for several samples and discussed a superposition of different types of defects contributing to the measured $1/f$-type noise. From the comparison of metallic and insulating samples we conclude that resistance noise in metallic Ga$_{1-x}$Mn$_{x}$As samples ($x>2$ \%) is mainly dominated by impurity switching processes and no prominent features occur around $T_\mathrm{C}$ even in the presence of an external out-of-plane magnetic field, while insulating samples, in particular Ga$_{1-x}$Mn$_{x}$P, show a sharp peak in the noise magnitude around $T_\mathrm{C}$ which can be attributed to percolation processes in the material, for which we find a scaling behavior $S_{R}/R^2 \propto R^{w}$. Consequently, for Ga$_{1-x}$Mn$_{x}$P, we infer that the picture of percolating magnetic polarons within the impurity-band model is applicable, while for Ga$_{1-x}$Mn$_{x}$As, this picture seems to be valid only for low Mn doping $x$. These findings for samples with localized charge carriers are supported by clear deviations between the calculated (within the DDH model) and experimentally determined frequency exponents $\alpha$ around the percolative transition. Besides varying the Mn content, another approach to tune Ga$_{1-x}$Mn$_{x}$As samples from the metallic side of the phase diagram towards the insulating regime is to irradiate the films with He ions. It was shown that fluctuation spectroscopy is sensitive to the changes in the defect landscape of irradiated samples, but no clear signs of electronic phase separation could be observed. \newline
 We have shown that a deeper understanding of defect physics and electronic phase separation in DMS can be obtained from fluctuation spectroscopy measurements. We suggest similar studies on other magnetic semiconductors which are supposed to exhibit percolation transitions or an electronic phase separation, like In$_{1-x}$Mn$_{x}$As or Mn$_{x}$Ge$_{1-x}$ \cite{Park2002}.

\section*{Acknowledgements}
The ion implantation was done at the Ion Beam Center (IBC) at HZDR. S.\ Z.\ acknowledges the financial support by the Helmholtz Association (VH-NG-713). We thank R.\ P.\ Campion from the University of Nottingham for supplying the Ga$_{1-x}$Mn$_{x}$As material for the He-ion irradiated samples.

%%%%%%%%%%%%%%%%%%%%%%%%%%%%%%%%%%%%%%%%%%%%%%%%%%%%%%%%%%%%%%%%%%%%%%%%%%%%%%%%%%%%%%%%%%%%%%%%%%%%%%%%%%%%%%%%%%%%%%%%%%%%%%%
\bibliography{GaMnAs_literature}

\begin{thebibliography}{61}%
\makeatletter
\providecommand \@ifxundefined [1]{%
 \@ifx{#1\undefined}
}%
\providecommand \@ifnum [1]{%
 \ifnum #1\expandafter \@firstoftwo
 \else \expandafter \@secondoftwo
 \fi
}%
\providecommand \@ifx [1]{%
 \ifx #1\expandafter \@firstoftwo
 \else \expandafter \@secondoftwo
 \fi
}%
\providecommand \natexlab [1]{#1}%
\providecommand \enquote  [1]{``#1''}%
\providecommand \bibnamefont  [1]{#1}%
\providecommand \bibfnamefont [1]{#1}%
\providecommand \citenamefont [1]{#1}%
\providecommand \href@noop [0]{\@secondoftwo}%
\providecommand \href [0]{\begingroup \@sanitize@url \@href}%
\providecommand \@href[1]{\@@startlink{#1}\@@href}%
\providecommand \@@href[1]{\endgroup#1\@@endlink}%
\providecommand \@sanitize@url [0]{\catcode `\\12\catcode `\$12\catcode
  `\&12\catcode `\#12\catcode `\^12\catcode `\_12\catcode `\%12\relax}%
\providecommand \@@startlink[1]{}%
\providecommand \@@endlink[0]{}%
\providecommand \url  [0]{\begingroup\@sanitize@url \@url }%
\providecommand \@url [1]{\endgroup\@href {#1}{\urlprefix }}%
\providecommand \urlprefix  [0]{URL }%
\providecommand \Eprint [0]{\href }%
\providecommand \doibase [0]{http://dx.doi.org/}%
\providecommand \selectlanguage [0]{\@gobble}%
\providecommand \bibinfo  [0]{\@secondoftwo}%
\providecommand \bibfield  [0]{\@secondoftwo}%
\providecommand \translation [1]{[#1]}%
\providecommand \BibitemOpen [0]{}%
\providecommand \bibitemStop [0]{}%
\providecommand \bibitemNoStop [0]{.\EOS\space}%
\providecommand \EOS [0]{\spacefactor3000\relax}%
\providecommand \BibitemShut  [1]{\csname bibitem#1\endcsname}%
\let\auto@bib@innerbib\@empty
%</preamble>
\bibitem [{\citenamefont {Dietl}\ and\ \citenamefont {Ohno}(2014)}]{Dietl2014}%
  \BibitemOpen
  \bibfield  {author} {\bibinfo {author} {\bibfnamefont {T.}~\bibnamefont
  {Dietl}}\ and\ \bibinfo {author} {\bibfnamefont {H.}~\bibnamefont {Ohno}},\
  }\href@noop {} {\bibfield  {journal} {\bibinfo  {journal} {Reviews of Modern
  Physics}\ }\textbf {\bibinfo {volume} {86}},\ \bibinfo {pages} {187}
  (\bibinfo {year} {2014})}\BibitemShut {NoStop}%
\bibitem [{\citenamefont {Dietl}(2010)}]{Dietl2010}%
  \BibitemOpen
  \bibfield  {author} {\bibinfo {author} {\bibfnamefont {T.}~\bibnamefont
  {Dietl}},\ }\href@noop {} {\bibfield  {journal} {\bibinfo  {journal} {Nature
  Materials}\ }\textbf {\bibinfo {volume} {9}},\ \bibinfo {pages} {965}
  (\bibinfo {year} {2010})}\BibitemShut {NoStop}%
\bibitem [{\citenamefont {Ohno}(1998)}]{Ohno1998}%
  \BibitemOpen
  \bibfield  {author} {\bibinfo {author} {\bibfnamefont {H.}~\bibnamefont
  {Ohno}},\ }\href@noop {} {\bibfield  {journal} {\bibinfo  {journal}
  {Science}\ }\textbf {\bibinfo {volume} {281}},\ \bibinfo {pages} {951}
  (\bibinfo {year} {1998})}\BibitemShut {NoStop}%
\bibitem [{\citenamefont {Jungwirth}\ \emph {et~al.}(2006)\citenamefont
  {Jungwirth}, \citenamefont {Sinova}, \citenamefont {Ma{\v{s}}ek},
  \citenamefont {Ku{\v{c}}era},\ and\ \citenamefont
  {MacDonald}}]{Jungwirth2006}%
  \BibitemOpen
  \bibfield  {author} {\bibinfo {author} {\bibfnamefont {T.}~\bibnamefont
  {Jungwirth}}, \bibinfo {author} {\bibfnamefont {J.}~\bibnamefont {Sinova}},
  \bibinfo {author} {\bibfnamefont {J.}~\bibnamefont {Ma{\v{s}}ek}}, \bibinfo
  {author} {\bibfnamefont {J.}~\bibnamefont {Ku{\v{c}}era}}, \ and\ \bibinfo
  {author} {\bibfnamefont {A.~H.}\ \bibnamefont {MacDonald}},\ }\href@noop {}
  {\bibfield  {journal} {\bibinfo  {journal} {Reviews of Modern Physics}\
  }\textbf {\bibinfo {volume} {78}},\ \bibinfo {pages} {809} (\bibinfo {year}
  {2006})}\BibitemShut {NoStop}%
\bibitem [{\citenamefont {Jungwirth}\ \emph {et~al.}(2007)\citenamefont
  {Jungwirth}, \citenamefont {Sinova}, \citenamefont {MacDonald}, \citenamefont
  {Gallagher}, \citenamefont {Nov{\'{a}}k}, \citenamefont {Edmonds},
  \citenamefont {Rushforth}, \citenamefont {Campion}, \citenamefont {Foxon},
  \citenamefont {Eaves}, \citenamefont {Olejn{\'{\i}}k}, \citenamefont
  {Ma{\v{s}}ek}, \citenamefont {Yang}, \citenamefont {Wunderlich},
  \citenamefont {Gould}, \citenamefont {Molenkamp}, \citenamefont {Dietl},\
  and\ \citenamefont {Ohno}}]{Jungwirth2007}%
  \BibitemOpen
  \bibfield  {author} {\bibinfo {author} {\bibfnamefont {T.}~\bibnamefont
  {Jungwirth}}, \bibinfo {author} {\bibfnamefont {J.}~\bibnamefont {Sinova}},
  \bibinfo {author} {\bibfnamefont {A.~H.}\ \bibnamefont {MacDonald}}, \bibinfo
  {author} {\bibfnamefont {B.~L.}\ \bibnamefont {Gallagher}}, \bibinfo {author}
  {\bibfnamefont {V.}~\bibnamefont {Nov{\'{a}}k}}, \bibinfo {author}
  {\bibfnamefont {K.~W.}\ \bibnamefont {Edmonds}}, \bibinfo {author}
  {\bibfnamefont {A.~W.}\ \bibnamefont {Rushforth}}, \bibinfo {author}
  {\bibfnamefont {R.~P.}\ \bibnamefont {Campion}}, \bibinfo {author}
  {\bibfnamefont {C.~T.}\ \bibnamefont {Foxon}}, \bibinfo {author}
  {\bibfnamefont {L.}~\bibnamefont {Eaves}}, \bibinfo {author} {\bibfnamefont
  {E.}~\bibnamefont {Olejn{\'{\i}}k}}, \bibinfo {author} {\bibfnamefont
  {J.}~\bibnamefont {Ma{\v{s}}ek}}, \bibinfo {author} {\bibfnamefont
  {S.-R.~E.}\ \bibnamefont {Yang}}, \bibinfo {author} {\bibfnamefont
  {J.}~\bibnamefont {Wunderlich}}, \bibinfo {author} {\bibfnamefont
  {C.}~\bibnamefont {Gould}}, \bibinfo {author} {\bibfnamefont {L.~W.}\
  \bibnamefont {Molenkamp}}, \bibinfo {author} {\bibfnamefont {T.}~\bibnamefont
  {Dietl}}, \ and\ \bibinfo {author} {\bibfnamefont {H.}~\bibnamefont {Ohno}},\
  }\href@noop {} {\bibfield  {journal} {\bibinfo  {journal} {Phys. Rev. B}\
  }\textbf {\bibinfo {volume} {76}},\ \bibinfo {pages} {125206} (\bibinfo
  {year} {2007})}\BibitemShut {NoStop}%
\bibitem [{\citenamefont {Scarpulla}\ \emph {et~al.}(2005)\citenamefont
  {Scarpulla}, \citenamefont {Cardozo}, \citenamefont {Farshchi}, \citenamefont
  {Oo}, \citenamefont {McCluskey}, \citenamefont {Yu},\ and\ \citenamefont
  {Dubon}}]{Scarpulla2005}%
  \BibitemOpen
  \bibfield  {author} {\bibinfo {author} {\bibfnamefont {M.~A.}\ \bibnamefont
  {Scarpulla}}, \bibinfo {author} {\bibfnamefont {B.~L.}\ \bibnamefont
  {Cardozo}}, \bibinfo {author} {\bibfnamefont {R.}~\bibnamefont {Farshchi}},
  \bibinfo {author} {\bibfnamefont {W.~M.~H.}\ \bibnamefont {Oo}}, \bibinfo
  {author} {\bibfnamefont {M.~D.}\ \bibnamefont {McCluskey}}, \bibinfo {author}
  {\bibfnamefont {K.~M.}\ \bibnamefont {Yu}}, \ and\ \bibinfo {author}
  {\bibfnamefont {O.~D.}\ \bibnamefont {Dubon}},\ }\href@noop {} {\bibfield
  {journal} {\bibinfo  {journal} {Phys. Rev. Lett.}\ }\textbf {\bibinfo
  {volume} {95}},\ \bibinfo {pages} {207204} (\bibinfo {year}
  {2005})}\BibitemShut {NoStop}%
\bibitem [{\citenamefont {Winkler}\ \emph {et~al.}(2011)\citenamefont
  {Winkler}, \citenamefont {Stone}, \citenamefont {Li}, \citenamefont {Yu},
  \citenamefont {Bonanni},\ and\ \citenamefont {Dubon}}]{Winkler2011}%
  \BibitemOpen
  \bibfield  {author} {\bibinfo {author} {\bibfnamefont {T.~E.}\ \bibnamefont
  {Winkler}}, \bibinfo {author} {\bibfnamefont {P.~R.}\ \bibnamefont {Stone}},
  \bibinfo {author} {\bibfnamefont {T.}~\bibnamefont {Li}}, \bibinfo {author}
  {\bibfnamefont {K.~M.}\ \bibnamefont {Yu}}, \bibinfo {author} {\bibfnamefont
  {A.}~\bibnamefont {Bonanni}}, \ and\ \bibinfo {author} {\bibfnamefont
  {O.~D.}\ \bibnamefont {Dubon}},\ }\href@noop {} {\bibfield  {journal}
  {\bibinfo  {journal} {Appl. Phys. Lett.}\ }\textbf {\bibinfo {volume} {98}},\
  \bibinfo {pages} {012103} (\bibinfo {year} {2011})}\BibitemShut {NoStop}%
\bibitem [{\citenamefont {Chen}\ \emph {et~al.}(2009)\citenamefont {Chen},
  \citenamefont {Yan}, \citenamefont {Xu}, \citenamefont {Lu}, \citenamefont
  {Wang}, \citenamefont {Deng}, \citenamefont {Qian}, \citenamefont {Ji},\ and\
  \citenamefont {Zhao}}]{Chen2009}%
  \BibitemOpen
  \bibfield  {author} {\bibinfo {author} {\bibfnamefont {L.}~\bibnamefont
  {Chen}}, \bibinfo {author} {\bibfnamefont {S.}~\bibnamefont {Yan}}, \bibinfo
  {author} {\bibfnamefont {P.~F.}\ \bibnamefont {Xu}}, \bibinfo {author}
  {\bibfnamefont {J.}~\bibnamefont {Lu}}, \bibinfo {author} {\bibfnamefont
  {W.~Z.}\ \bibnamefont {Wang}}, \bibinfo {author} {\bibfnamefont {J.~J.}\
  \bibnamefont {Deng}}, \bibinfo {author} {\bibfnamefont {X.}~\bibnamefont
  {Qian}}, \bibinfo {author} {\bibfnamefont {Y.}~\bibnamefont {Ji}}, \ and\
  \bibinfo {author} {\bibfnamefont {J.~H.}\ \bibnamefont {Zhao}},\ }\href@noop
  {} {\bibfield  {journal} {\bibinfo  {journal} {Appl. Phys. Lett.}\ }\textbf
  {\bibinfo {volume} {95}},\ \bibinfo {pages} {182505} (\bibinfo {year}
  {2009})}\BibitemShut {NoStop}%
\bibitem [{\citenamefont {Farshchi}\ \emph {et~al.}(2006)\citenamefont
  {Farshchi}, \citenamefont {Scarpulla}, \citenamefont {Stone}, \citenamefont
  {Yu}, \citenamefont {Sharp}, \citenamefont {Beeman}, \citenamefont
  {Silvestri}, \citenamefont {Reichertz}, \citenamefont {Haller},\ and\
  \citenamefont {Dubon}}]{Farshchi2006}%
  \BibitemOpen
  \bibfield  {author} {\bibinfo {author} {\bibfnamefont {R.}~\bibnamefont
  {Farshchi}}, \bibinfo {author} {\bibfnamefont {M.}~\bibnamefont {Scarpulla}},
  \bibinfo {author} {\bibfnamefont {P.}~\bibnamefont {Stone}}, \bibinfo
  {author} {\bibfnamefont {K.}~\bibnamefont {Yu}}, \bibinfo {author}
  {\bibfnamefont {I.}~\bibnamefont {Sharp}}, \bibinfo {author} {\bibfnamefont
  {J.}~\bibnamefont {Beeman}}, \bibinfo {author} {\bibfnamefont
  {H.}~\bibnamefont {Silvestri}}, \bibinfo {author} {\bibfnamefont
  {L.}~\bibnamefont {Reichertz}}, \bibinfo {author} {\bibfnamefont
  {E.}~\bibnamefont {Haller}}, \ and\ \bibinfo {author} {\bibfnamefont
  {O.}~\bibnamefont {Dubon}},\ }\href@noop {} {\bibfield  {journal} {\bibinfo
  {journal} {Solid State Communications}\ }\textbf {\bibinfo {volume} {140}},\
  \bibinfo {pages} {443} (\bibinfo {year} {2006})}\BibitemShut {NoStop}%
\bibitem [{\citenamefont {Berciu}\ and\ \citenamefont
  {Bhatt}(2001)}]{Berciu2001}%
  \BibitemOpen
  \bibfield  {author} {\bibinfo {author} {\bibfnamefont {M.}~\bibnamefont
  {Berciu}}\ and\ \bibinfo {author} {\bibfnamefont {R.~N.}\ \bibnamefont
  {Bhatt}},\ }\href@noop {} {\bibfield  {journal} {\bibinfo  {journal} {Phys.
  Rev. Lett.}\ }\textbf {\bibinfo {volume} {87}},\ \bibinfo {pages} {107203}
  (\bibinfo {year} {2001})}\BibitemShut {NoStop}%
\bibitem [{\citenamefont {Richardella}\ \emph {et~al.}(2010)\citenamefont
  {Richardella}, \citenamefont {Roushan}, \citenamefont {Mack}, \citenamefont
  {Zhou}, \citenamefont {Huse}, \citenamefont {Awschalom},\ and\ \citenamefont
  {Yazdani}}]{Richardella2010}%
  \BibitemOpen
  \bibfield  {author} {\bibinfo {author} {\bibfnamefont {A.}~\bibnamefont
  {Richardella}}, \bibinfo {author} {\bibfnamefont {P.}~\bibnamefont
  {Roushan}}, \bibinfo {author} {\bibfnamefont {S.}~\bibnamefont {Mack}},
  \bibinfo {author} {\bibfnamefont {B.}~\bibnamefont {Zhou}}, \bibinfo {author}
  {\bibfnamefont {D.~A.}\ \bibnamefont {Huse}}, \bibinfo {author}
  {\bibfnamefont {D.~D.}\ \bibnamefont {Awschalom}}, \ and\ \bibinfo {author}
  {\bibfnamefont {A.}~\bibnamefont {Yazdani}},\ }\href@noop {} {\bibfield
  {journal} {\bibinfo  {journal} {Science}\ }\textbf {\bibinfo {volume}
  {327}},\ \bibinfo {pages} {665} (\bibinfo {year} {2010})}\BibitemShut
  {NoStop}%
\bibitem [{\citenamefont {Sato}\ \emph {et~al.}(2010)\citenamefont {Sato},
  \citenamefont {Bergqvist}, \citenamefont {Kudrnovsk{\'{y}}}, \citenamefont
  {Dederichs}, \citenamefont {Eriksson}, \citenamefont {Turek}, \citenamefont
  {Sanyal}, \citenamefont {Bouzerar}, \citenamefont {Katayama-Yoshida},
  \citenamefont {Dinh}, \citenamefont {Fukushima}, \citenamefont {Kizaki},\
  and\ \citenamefont {Zeller}}]{Sato2010}%
  \BibitemOpen
  \bibfield  {author} {\bibinfo {author} {\bibfnamefont {K.}~\bibnamefont
  {Sato}}, \bibinfo {author} {\bibfnamefont {L.}~\bibnamefont {Bergqvist}},
  \bibinfo {author} {\bibfnamefont {J.}~\bibnamefont {Kudrnovsk{\'{y}}}},
  \bibinfo {author} {\bibfnamefont {P.~H.}\ \bibnamefont {Dederichs}}, \bibinfo
  {author} {\bibfnamefont {O.}~\bibnamefont {Eriksson}}, \bibinfo {author}
  {\bibfnamefont {I.}~\bibnamefont {Turek}}, \bibinfo {author} {\bibfnamefont
  {B.}~\bibnamefont {Sanyal}}, \bibinfo {author} {\bibfnamefont
  {G.}~\bibnamefont {Bouzerar}}, \bibinfo {author} {\bibfnamefont
  {H.}~\bibnamefont {Katayama-Yoshida}}, \bibinfo {author} {\bibfnamefont
  {V.~A.}\ \bibnamefont {Dinh}}, \bibinfo {author} {\bibfnamefont
  {T.}~\bibnamefont {Fukushima}}, \bibinfo {author} {\bibfnamefont
  {H.}~\bibnamefont {Kizaki}}, \ and\ \bibinfo {author} {\bibfnamefont
  {R.}~\bibnamefont {Zeller}},\ }\href@noop {} {\bibfield  {journal} {\bibinfo
  {journal} {Reviews of Modern Physics}\ }\textbf {\bibinfo {volume} {82}},\
  \bibinfo {pages} {1633} (\bibinfo {year} {2010})}\BibitemShut {NoStop}%
\bibitem [{\citenamefont {Samarth}(2012)}]{Samarth2012}%
  \BibitemOpen
  \bibfield  {author} {\bibinfo {author} {\bibfnamefont {N.}~\bibnamefont
  {Samarth}},\ }\href@noop {} {\bibfield  {journal} {\bibinfo  {journal}
  {Nature Materials}\ }\textbf {\bibinfo {volume} {11}},\ \bibinfo {pages}
  {360} (\bibinfo {year} {2012})}\BibitemShut {NoStop}%
\bibitem [{\citenamefont {K\"{o}nig}\ \emph {et~al.}(2003)\citenamefont
  {K\"{o}nig}, \citenamefont {Schliemann}, \citenamefont {Jungwirth},\ and\
  \citenamefont {MacDonald}}]{Koenig2003}%
  \BibitemOpen
  \bibfield  {author} {\bibinfo {author} {\bibfnamefont {J.}~\bibnamefont
  {K\"{o}nig}}, \bibinfo {author} {\bibfnamefont {J.}~\bibnamefont
  {Schliemann}}, \bibinfo {author} {\bibfnamefont {T.}~\bibnamefont
  {Jungwirth}}, \ and\ \bibinfo {author} {\bibfnamefont {A.~H.}\ \bibnamefont
  {MacDonald}},\ }in\ \href@noop {} {\emph {\bibinfo {booktitle} {Electronic
  Structure and Magnetism of Complex Materials}}}\ (\bibinfo  {publisher}
  {Springer Science $+$ Business Media},\ \bibinfo {year} {2003})\ pp.\
  \bibinfo {pages} {163--211}\BibitemShut {NoStop}%
\bibitem [{\citenamefont {Dietl}\ \emph {et~al.}(2001)\citenamefont {Dietl},
  \citenamefont {Ohno},\ and\ \citenamefont {Matsukura}}]{Dietl2001}%
  \BibitemOpen
  \bibfield  {author} {\bibinfo {author} {\bibfnamefont {T.}~\bibnamefont
  {Dietl}}, \bibinfo {author} {\bibfnamefont {H.}~\bibnamefont {Ohno}}, \ and\
  \bibinfo {author} {\bibfnamefont {F.}~\bibnamefont {Matsukura}},\ }\href@noop
  {} {\bibfield  {journal} {\bibinfo  {journal} {Phys. Rev. B}\ }\textbf
  {\bibinfo {volume} {63}},\ \bibinfo {pages} {195205} (\bibinfo {year}
  {2001})}\BibitemShut {NoStop}%
\bibitem [{\citenamefont {Dietl}\ \emph {et~al.}(2000)\citenamefont {Dietl},
  \citenamefont {Ohno}, \citenamefont {Matsukura}, \citenamefont {Cibert},\
  and\ \citenamefont {Ferrand}}]{Dietl2000}%
  \BibitemOpen
  \bibfield  {author} {\bibinfo {author} {\bibfnamefont {T.}~\bibnamefont
  {Dietl}}, \bibinfo {author} {\bibfnamefont {H.}~\bibnamefont {Ohno}},
  \bibinfo {author} {\bibfnamefont {F.}~\bibnamefont {Matsukura}}, \bibinfo
  {author} {\bibfnamefont {J.}~\bibnamefont {Cibert}}, \ and\ \bibinfo {author}
  {\bibfnamefont {D.}~\bibnamefont {Ferrand}},\ }\href@noop {} {\bibfield
  {journal} {\bibinfo  {journal} {Science}\ }\textbf {\bibinfo {volume}
  {287}},\ \bibinfo {pages} {1019} (\bibinfo {year} {2000})}\BibitemShut
  {NoStop}%
\bibitem [{\citenamefont {K\"{o}nig}\ \emph {et~al.}(2000)\citenamefont
  {K\"{o}nig}, \citenamefont {Lin},\ and\ \citenamefont
  {MacDonald}}]{Koenig2000}%
  \BibitemOpen
  \bibfield  {author} {\bibinfo {author} {\bibfnamefont {J.}~\bibnamefont
  {K\"{o}nig}}, \bibinfo {author} {\bibfnamefont {H.-H.}\ \bibnamefont {Lin}},
  \ and\ \bibinfo {author} {\bibfnamefont {A.~H.}\ \bibnamefont {MacDonald}},\
  }\href@noop {} {\bibfield  {journal} {\bibinfo  {journal} {Phys. Rev. Lett.}\
  }\textbf {\bibinfo {volume} {84}},\ \bibinfo {pages} {5628} (\bibinfo {year}
  {2000})}\BibitemShut {NoStop}%
\bibitem [{\citenamefont {Dobrowolska}\ \emph {et~al.}(2012)\citenamefont
  {Dobrowolska}, \citenamefont {Tivakornsasithorn}, \citenamefont {Liu},
  \citenamefont {Furdyna}, \citenamefont {Berciu}, \citenamefont {Yu},\ and\
  \citenamefont {Walukiewicz}}]{Dobrowolska2012}%
  \BibitemOpen
  \bibfield  {author} {\bibinfo {author} {\bibfnamefont {M.}~\bibnamefont
  {Dobrowolska}}, \bibinfo {author} {\bibfnamefont {K.}~\bibnamefont
  {Tivakornsasithorn}}, \bibinfo {author} {\bibfnamefont {X.}~\bibnamefont
  {Liu}}, \bibinfo {author} {\bibfnamefont {J.~K.}\ \bibnamefont {Furdyna}},
  \bibinfo {author} {\bibfnamefont {M.}~\bibnamefont {Berciu}}, \bibinfo
  {author} {\bibfnamefont {K.~M.}\ \bibnamefont {Yu}}, \ and\ \bibinfo {author}
  {\bibfnamefont {W.}~\bibnamefont {Walukiewicz}},\ }\href@noop {} {\bibfield
  {journal} {\bibinfo  {journal} {Nature Materials}\ }\textbf {\bibinfo
  {volume} {11}},\ \bibinfo {pages} {444} (\bibinfo {year} {2012})}\BibitemShut
  {NoStop}%
\bibitem [{\citenamefont {Bhatt}\ \emph {et~al.}(2002)\citenamefont {Bhatt},
  \citenamefont {Berciu}, \citenamefont {Kennett},\ and\ \citenamefont
  {Wan}}]{Bhatt2002}%
  \BibitemOpen
  \bibfield  {author} {\bibinfo {author} {\bibfnamefont {R.~N.}\ \bibnamefont
  {Bhatt}}, \bibinfo {author} {\bibfnamefont {M.}~\bibnamefont {Berciu}},
  \bibinfo {author} {\bibfnamefont {M.~P.}\ \bibnamefont {Kennett}}, \ and\
  \bibinfo {author} {\bibfnamefont {X.}~\bibnamefont {Wan}},\ }\href@noop {}
  {\bibfield  {journal} {\bibinfo  {journal} {Journal of Superconductivity:
  Incorporating Novel Magnetism}\ }\textbf {\bibinfo {volume} {15}},\ \bibinfo
  {pages} {71} (\bibinfo {year} {2002})}\BibitemShut {NoStop}%
\bibitem [{\citenamefont {Kaminski}\ and\ \citenamefont
  {Sarma}(2002)}]{Kaminski2002}%
  \BibitemOpen
  \bibfield  {author} {\bibinfo {author} {\bibfnamefont {A.}~\bibnamefont
  {Kaminski}}\ and\ \bibinfo {author} {\bibfnamefont {S.~D.}\ \bibnamefont
  {Sarma}},\ }\href@noop {} {\bibfield  {journal} {\bibinfo  {journal} {Phys.
  Rev. Lett.}\ }\textbf {\bibinfo {volume} {88}},\ \bibinfo {pages} {247202}
  (\bibinfo {year} {2002})}\BibitemShut {NoStop}%
\bibitem [{\citenamefont {Kaminski}\ and\ \citenamefont
  {Sarma}(2003)}]{Kaminski2003}%
  \BibitemOpen
  \bibfield  {author} {\bibinfo {author} {\bibfnamefont {A.}~\bibnamefont
  {Kaminski}}\ and\ \bibinfo {author} {\bibfnamefont {S.~D.}\ \bibnamefont
  {Sarma}},\ }\href@noop {} {\bibfield  {journal} {\bibinfo  {journal} {Phys.
  Rev. B}\ }\textbf {\bibinfo {volume} {68}},\ \bibinfo {pages} {235210}
  (\bibinfo {year} {2003})}\BibitemShut {NoStop}%
\bibitem [{\citenamefont {Kumar}\ \emph {et~al.}(2013)\citenamefont {Kumar},
  \citenamefont {Paschoal}, \citenamefont {Johannes}, \citenamefont
  {Jacobsson}, \citenamefont {Borschel}, \citenamefont {Pertsova},
  \citenamefont {Wang}, \citenamefont {Wu}, \citenamefont {Canali},
  \citenamefont {Ronning}, \citenamefont {Samuelson},\ and\ \citenamefont
  {Pettersson}}]{Kumar2013}%
  \BibitemOpen
  \bibfield  {author} {\bibinfo {author} {\bibfnamefont {S.}~\bibnamefont
  {Kumar}}, \bibinfo {author} {\bibfnamefont {W.}~\bibnamefont {Paschoal}},
  \bibinfo {author} {\bibfnamefont {A.}~\bibnamefont {Johannes}}, \bibinfo
  {author} {\bibfnamefont {D.}~\bibnamefont {Jacobsson}}, \bibinfo {author}
  {\bibfnamefont {C.}~\bibnamefont {Borschel}}, \bibinfo {author}
  {\bibfnamefont {A.}~\bibnamefont {Pertsova}}, \bibinfo {author}
  {\bibfnamefont {C.-H.}\ \bibnamefont {Wang}}, \bibinfo {author}
  {\bibfnamefont {M.-K.}\ \bibnamefont {Wu}}, \bibinfo {author} {\bibfnamefont
  {C.~M.}\ \bibnamefont {Canali}}, \bibinfo {author} {\bibfnamefont
  {C.}~\bibnamefont {Ronning}}, \bibinfo {author} {\bibfnamefont
  {L.}~\bibnamefont {Samuelson}}, \ and\ \bibinfo {author} {\bibfnamefont
  {H.}~\bibnamefont {Pettersson}},\ }\href@noop {} {\bibfield  {journal}
  {\bibinfo  {journal} {Nano Letters}\ }\textbf {\bibinfo {volume} {13}},\
  \bibinfo {pages} {5079} (\bibinfo {year} {2013})}\BibitemShut {NoStop}%
\bibitem [{\citenamefont {Ohno}\ and\ \citenamefont {Dietl}(2008)}]{Ohno2008}%
  \BibitemOpen
  \bibfield  {author} {\bibinfo {author} {\bibfnamefont {H.}~\bibnamefont
  {Ohno}}\ and\ \bibinfo {author} {\bibfnamefont {T.}~\bibnamefont {Dietl}},\
  }\href@noop {} {\bibfield  {journal} {\bibinfo  {journal} {Journal of
  Magnetism and Magnetic Materials}\ }\textbf {\bibinfo {volume} {320}},\
  \bibinfo {pages} {1293} (\bibinfo {year} {2008})}\BibitemShut {NoStop}%
\bibitem [{\citenamefont {Dietl}(2006)}]{Dietl2006}%
  \BibitemOpen
  \bibfield  {author} {\bibinfo {author} {\bibfnamefont {T.}~\bibnamefont
  {Dietl}},\ }\href@noop {} {\bibfield  {journal} {\bibinfo  {journal} {Physica
  E: Low-dimensional Systems and Nanostructures}\ }\textbf {\bibinfo {volume}
  {35}},\ \bibinfo {pages} {293} (\bibinfo {year} {2006})}\BibitemShut
  {NoStop}%
\bibitem [{\citenamefont {Podzorov}\ \emph {et~al.}(2000)\citenamefont
  {Podzorov}, \citenamefont {Uehara}, \citenamefont {Gershenson}, \citenamefont
  {Koo},\ and\ \citenamefont {Cheong}}]{Podzorov2000}%
  \BibitemOpen
  \bibfield  {author} {\bibinfo {author} {\bibfnamefont {V.}~\bibnamefont
  {Podzorov}}, \bibinfo {author} {\bibfnamefont {M.}~\bibnamefont {Uehara}},
  \bibinfo {author} {\bibfnamefont {M.~E.}\ \bibnamefont {Gershenson}},
  \bibinfo {author} {\bibfnamefont {T.~Y.}\ \bibnamefont {Koo}}, \ and\
  \bibinfo {author} {\bibfnamefont {S.-W.}\ \bibnamefont {Cheong}},\
  }\href@noop {} {\bibfield  {journal} {\bibinfo  {journal} {Phys. Rev. B}\
  }\textbf {\bibinfo {volume} {61}},\ \bibinfo {pages} {R3784} (\bibinfo {year}
  {2000})}\BibitemShut {NoStop}%
\bibitem [{\citenamefont {Das}\ \emph {et~al.}(2012)\citenamefont {Das},
  \citenamefont {Amyan}, \citenamefont {Brandenburg}, \citenamefont {Mueller},
  \citenamefont {Xiong}, \citenamefont {von Moln{\'{a}}r},\ and\ \citenamefont
  {Fisk}}]{Das2012}%
  \BibitemOpen
  \bibfield  {author} {\bibinfo {author} {\bibfnamefont {P.}~\bibnamefont
  {Das}}, \bibinfo {author} {\bibfnamefont {A.}~\bibnamefont {Amyan}}, \bibinfo
  {author} {\bibfnamefont {J.}~\bibnamefont {Brandenburg}}, \bibinfo {author}
  {\bibfnamefont {J.}~\bibnamefont {Mueller}}, \bibinfo {author} {\bibfnamefont
  {P.}~\bibnamefont {Xiong}}, \bibinfo {author} {\bibfnamefont
  {S.}~\bibnamefont {von Moln{\'{a}}r}}, \ and\ \bibinfo {author}
  {\bibfnamefont {Z.}~\bibnamefont {Fisk}},\ }\href@noop {} {\bibfield
  {journal} {\bibinfo  {journal} {Phys. Rev. B}\ }\textbf {\bibinfo {volume}
  {86}},\ \bibinfo {pages} {184425} (\bibinfo {year} {2012})}\BibitemShut
  {NoStop}%
\bibitem [{\citenamefont {Clerjaud}(1985)}]{Clerjaud1985}%
  \BibitemOpen
  \bibfield  {author} {\bibinfo {author} {\bibfnamefont {B.}~\bibnamefont
  {Clerjaud}},\ }\href@noop {} {\bibfield  {journal} {\bibinfo  {journal}
  {Journal of Physics C: Solid State Physics}\ }\textbf {\bibinfo {volume}
  {18}},\ \bibinfo {pages} {3615} (\bibinfo {year} {1985})}\BibitemShut
  {NoStop}%
\bibitem [{\citenamefont {Zhou}(2015)}]{Zhou2015}%
  \BibitemOpen
  \bibfield  {author} {\bibinfo {author} {\bibfnamefont {S.}~\bibnamefont
  {Zhou}},\ }\href@noop {} {\bibfield  {journal} {\bibinfo  {journal} {Journal
  of Physics D: Applied Physics}\ }\textbf {\bibinfo {volume} {48}},\ \bibinfo
  {pages} {263001} (\bibinfo {year} {2015})}\BibitemShut {NoStop}%
\bibitem [{\citenamefont {Lonsky}\ \emph {et~al.}(2016)\citenamefont {Lonsky},
  \citenamefont {Heinz}, \citenamefont {Daniel}, \citenamefont {Albrecht},\
  and\ \citenamefont {M\"{u}ller}}]{Lonsky2016}%
  \BibitemOpen
  \bibfield  {author} {\bibinfo {author} {\bibfnamefont {M.}~\bibnamefont
  {Lonsky}}, \bibinfo {author} {\bibfnamefont {S.}~\bibnamefont {Heinz}},
  \bibinfo {author} {\bibfnamefont {M.~V.}\ \bibnamefont {Daniel}}, \bibinfo
  {author} {\bibfnamefont {M.}~\bibnamefont {Albrecht}}, \ and\ \bibinfo
  {author} {\bibfnamefont {J.}~\bibnamefont {M\"{u}ller}},\ }\href@noop {}
  {\bibfield  {journal} {\bibinfo  {journal} {J. Appl. Phys.}\ }\textbf
  {\bibinfo {volume} {120}},\ \bibinfo {pages} {142101} (\bibinfo {year}
  {2016})}\BibitemShut {NoStop}%
\bibitem [{\citenamefont {Lonsky}\ \emph {et~al.}(2015)\citenamefont {Lonsky},
  \citenamefont {Heinz}, \citenamefont {Daniel}, \citenamefont {Albrecht},\
  and\ \citenamefont {M\"{u}ller}}]{Lonsky2015}%
  \BibitemOpen
  \bibfield  {author} {\bibinfo {author} {\bibfnamefont {M.}~\bibnamefont
  {Lonsky}}, \bibinfo {author} {\bibfnamefont {S.}~\bibnamefont {Heinz}},
  \bibinfo {author} {\bibfnamefont {M.}~\bibnamefont {Daniel}}, \bibinfo
  {author} {\bibfnamefont {M.}~\bibnamefont {Albrecht}}, \ and\ \bibinfo
  {author} {\bibfnamefont {J.}~\bibnamefont {M\"{u}ller}},\ }in\ \href@noop {}
  {\emph {\bibinfo {booktitle} {2015 International Conference on Noise and
  Fluctuations ({ICNF})}}}\ (\bibinfo  {publisher} {Institute of Electrical and
  Electronics Engineers ({IEEE})},\ \bibinfo {year} {2015})\BibitemShut
  {NoStop}%
\bibitem [{\citenamefont {M\"{u}ller}\ \emph
  {et~al.}(2006{\natexlab{a}})\citenamefont {M\"{u}ller}, \citenamefont {von
  Moln{\'{a}}r}, \citenamefont {Ohno},\ and\ \citenamefont
  {Ohno}}]{Mueller2006}%
  \BibitemOpen
  \bibfield  {author} {\bibinfo {author} {\bibfnamefont {J.}~\bibnamefont
  {M\"{u}ller}}, \bibinfo {author} {\bibfnamefont {S.}~\bibnamefont {von
  Moln{\'{a}}r}}, \bibinfo {author} {\bibfnamefont {Y.}~\bibnamefont {Ohno}}, \
  and\ \bibinfo {author} {\bibfnamefont {H.}~\bibnamefont {Ohno}},\ }\href@noop
  {} {\bibfield  {journal} {\bibinfo  {journal} {Phys. Rev. Lett.}\ }\textbf
  {\bibinfo {volume} {96}},\ \bibinfo {pages} {186601} (\bibinfo {year}
  {2006}{\natexlab{a}})}\BibitemShut {NoStop}%
\bibitem [{\citenamefont {M\"{u}ller}\ \emph
  {et~al.}(2006{\natexlab{b}})\citenamefont {M\"{u}ller}, \citenamefont {Li},
  \citenamefont {von Moln{\'{a}}r}, \citenamefont {Ohno},\ and\ \citenamefont
  {Ohno}}]{Mueller2006a}%
  \BibitemOpen
  \bibfield  {author} {\bibinfo {author} {\bibfnamefont {J.}~\bibnamefont
  {M\"{u}ller}}, \bibinfo {author} {\bibfnamefont {Y.}~\bibnamefont {Li}},
  \bibinfo {author} {\bibfnamefont {S.}~\bibnamefont {von Moln{\'{a}}r}},
  \bibinfo {author} {\bibfnamefont {Y.}~\bibnamefont {Ohno}}, \ and\ \bibinfo
  {author} {\bibfnamefont {H.}~\bibnamefont {Ohno}},\ }\href@noop {} {\bibfield
   {journal} {\bibinfo  {journal} {Phys. Rev. B}\ }\textbf {\bibinfo {volume}
  {74}},\ \bibinfo {pages} {125310} (\bibinfo {year}
  {2006}{\natexlab{b}})}\BibitemShut {NoStop}%
\bibitem [{\citenamefont {M\"{u}ller}\ \emph {et~al.}(2015)\citenamefont
  {M\"{u}ller}, \citenamefont {K\"{o}rbitzer}, \citenamefont {Amyan},
  \citenamefont {Pohlit}, \citenamefont {Ohno},\ and\ \citenamefont
  {Ohno}}]{Muller2015}%
  \BibitemOpen
  \bibfield  {author} {\bibinfo {author} {\bibfnamefont {J.}~\bibnamefont
  {M\"{u}ller}}, \bibinfo {author} {\bibfnamefont {B.}~\bibnamefont
  {K\"{o}rbitzer}}, \bibinfo {author} {\bibfnamefont {A.}~\bibnamefont
  {Amyan}}, \bibinfo {author} {\bibfnamefont {M.}~\bibnamefont {Pohlit}},
  \bibinfo {author} {\bibfnamefont {Y.}~\bibnamefont {Ohno}}, \ and\ \bibinfo
  {author} {\bibfnamefont {H.}~\bibnamefont {Ohno}},\ }in\ \href@noop {} {\emph
  {\bibinfo {booktitle} {2015 International Conference on Noise and
  Fluctuations ({ICNF})}}}\ (\bibinfo  {publisher} {{IEEE}},\ \bibinfo {year}
  {2015})\BibitemShut {NoStop}%
\bibitem [{\citenamefont {Zhou}\ \emph {et~al.}(2016)\citenamefont {Zhou},
  \citenamefont {Li}, \citenamefont {Yuan}, \citenamefont {Rushforth},
  \citenamefont {Chen}, \citenamefont {Wang}, \citenamefont {B\"{o}ttger},
  \citenamefont {Heller}, \citenamefont {Zhao}, \citenamefont {Edmonds},
  \citenamefont {Campion}, \citenamefont {Gallagher}, \citenamefont {Timm},\
  and\ \citenamefont {Helm}}]{Zhou2016}%
  \BibitemOpen
  \bibfield  {author} {\bibinfo {author} {\bibfnamefont {S.}~\bibnamefont
  {Zhou}}, \bibinfo {author} {\bibfnamefont {L.}~\bibnamefont {Li}}, \bibinfo
  {author} {\bibfnamefont {Y.}~\bibnamefont {Yuan}}, \bibinfo {author}
  {\bibfnamefont {A.~W.}\ \bibnamefont {Rushforth}}, \bibinfo {author}
  {\bibfnamefont {L.}~\bibnamefont {Chen}}, \bibinfo {author} {\bibfnamefont
  {Y.}~\bibnamefont {Wang}}, \bibinfo {author} {\bibfnamefont {R.}~\bibnamefont
  {B\"{o}ttger}}, \bibinfo {author} {\bibfnamefont {R.}~\bibnamefont {Heller}},
  \bibinfo {author} {\bibfnamefont {J.}~\bibnamefont {Zhao}}, \bibinfo {author}
  {\bibfnamefont {K.~W.}\ \bibnamefont {Edmonds}}, \bibinfo {author}
  {\bibfnamefont {R.~P.}\ \bibnamefont {Campion}}, \bibinfo {author}
  {\bibfnamefont {B.~L.}\ \bibnamefont {Gallagher}}, \bibinfo {author}
  {\bibfnamefont {C.}~\bibnamefont {Timm}}, \ and\ \bibinfo {author}
  {\bibfnamefont {M.}~\bibnamefont {Helm}},\ }\href@noop {} {\bibfield
  {journal} {\bibinfo  {journal} {Phys. Rev. B}\ }\textbf {\bibinfo {volume}
  {94}},\ \bibinfo {pages} {075205} (\bibinfo {year} {2016})}\BibitemShut
  {NoStop}%
\bibitem [{\citenamefont {Hamida}\ \emph {et~al.}(2014)\citenamefont {Hamida},
  \citenamefont {Sievers}, \citenamefont {Bergmann}, \citenamefont {Racu},
  \citenamefont {Bieler}, \citenamefont {Pierz},\ and\ \citenamefont
  {Schumacher}}]{Hamida2014}%
  \BibitemOpen
  \bibfield  {author} {\bibinfo {author} {\bibfnamefont {A.~B.}\ \bibnamefont
  {Hamida}}, \bibinfo {author} {\bibfnamefont {S.}~\bibnamefont {Sievers}},
  \bibinfo {author} {\bibfnamefont {F.}~\bibnamefont {Bergmann}}, \bibinfo
  {author} {\bibfnamefont {A.}~\bibnamefont {Racu}}, \bibinfo {author}
  {\bibfnamefont {M.}~\bibnamefont {Bieler}}, \bibinfo {author} {\bibfnamefont
  {K.}~\bibnamefont {Pierz}}, \ and\ \bibinfo {author} {\bibfnamefont {H.~W.}\
  \bibnamefont {Schumacher}},\ }\href@noop {} {\bibfield  {journal} {\bibinfo
  {journal} {Phys. Status Solidi B}\ }\textbf {\bibinfo {volume} {251}},\
  \bibinfo {pages} {1652} (\bibinfo {year} {2014})}\BibitemShut {NoStop}%
\bibitem [{\citenamefont {Yuan}\ \emph {et~al.}(2017)\citenamefont {Yuan},
  \citenamefont {Xu}, \citenamefont {H\"{u}bner}, \citenamefont {Jakiela},
  \citenamefont {B\"{o}ttger}, \citenamefont {Helm}, \citenamefont {Sawicki},
  \citenamefont {Dietl},\ and\ \citenamefont {Zhou}}]{Yuan2017a}%
  \BibitemOpen
  \bibfield  {author} {\bibinfo {author} {\bibfnamefont {Y.}~\bibnamefont
  {Yuan}}, \bibinfo {author} {\bibfnamefont {C.}~\bibnamefont {Xu}}, \bibinfo
  {author} {\bibfnamefont {R.}~\bibnamefont {H\"{u}bner}}, \bibinfo {author}
  {\bibfnamefont {R.}~\bibnamefont {Jakiela}}, \bibinfo {author} {\bibfnamefont
  {R.}~\bibnamefont {B\"{o}ttger}}, \bibinfo {author} {\bibfnamefont
  {M.}~\bibnamefont {Helm}}, \bibinfo {author} {\bibfnamefont {M.}~\bibnamefont
  {Sawicki}}, \bibinfo {author} {\bibfnamefont {T.}~\bibnamefont {Dietl}}, \
  and\ \bibinfo {author} {\bibfnamefont {S.}~\bibnamefont {Zhou}},\ }\href@noop
  {} {\bibfield  {journal} {\bibinfo  {journal} {Physical Review Materials}\
  }\textbf {\bibinfo {volume} {1}},\ \bibinfo {pages} {054401} (\bibinfo {year}
  {2017})}\BibitemShut {NoStop}%
\bibitem [{\citenamefont {Hayashi}\ \emph {et~al.}(2001)\citenamefont
  {Hayashi}, \citenamefont {Hashimoto}, \citenamefont {Katsumoto},\ and\
  \citenamefont {Iye}}]{Hayashi2001}%
  \BibitemOpen
  \bibfield  {author} {\bibinfo {author} {\bibfnamefont {T.}~\bibnamefont
  {Hayashi}}, \bibinfo {author} {\bibfnamefont {Y.}~\bibnamefont {Hashimoto}},
  \bibinfo {author} {\bibfnamefont {S.}~\bibnamefont {Katsumoto}}, \ and\
  \bibinfo {author} {\bibfnamefont {Y.}~\bibnamefont {Iye}},\ }\href@noop {}
  {\bibfield  {journal} {\bibinfo  {journal} {Appl. Phys. Lett.}\ }\textbf
  {\bibinfo {volume} {78}},\ \bibinfo {pages} {1691} (\bibinfo {year}
  {2001})}\BibitemShut {NoStop}%
\bibitem [{\citenamefont {Potashnik}\ \emph {et~al.}(2001)\citenamefont
  {Potashnik}, \citenamefont {Ku}, \citenamefont {Chun}, \citenamefont {Berry},
  \citenamefont {Samarth},\ and\ \citenamefont {Schiffer}}]{Potashnik2001}%
  \BibitemOpen
  \bibfield  {author} {\bibinfo {author} {\bibfnamefont {S.~J.}\ \bibnamefont
  {Potashnik}}, \bibinfo {author} {\bibfnamefont {K.~C.}\ \bibnamefont {Ku}},
  \bibinfo {author} {\bibfnamefont {S.~H.}\ \bibnamefont {Chun}}, \bibinfo
  {author} {\bibfnamefont {J.~J.}\ \bibnamefont {Berry}}, \bibinfo {author}
  {\bibfnamefont {N.}~\bibnamefont {Samarth}}, \ and\ \bibinfo {author}
  {\bibfnamefont {P.}~\bibnamefont {Schiffer}},\ }\href@noop {} {\bibfield
  {journal} {\bibinfo  {journal} {Appl. Phys. Lett.}\ }\textbf {\bibinfo
  {volume} {79}},\ \bibinfo {pages} {1495} (\bibinfo {year}
  {2001})}\BibitemShut {NoStop}%
\bibitem [{\citenamefont {Esch}\ \emph {et~al.}(1997)\citenamefont {Esch},
  \citenamefont {Bockstal}, \citenamefont {Boeck}, \citenamefont {Verbanck},
  \citenamefont {van Steenbergen}, \citenamefont {Wellmann}, \citenamefont
  {Grietens}, \citenamefont {Bogaerts}, \citenamefont {Herlach},\ and\
  \citenamefont {Borghs}}]{Esch1997}%
  \BibitemOpen
  \bibfield  {author} {\bibinfo {author} {\bibfnamefont {A.~V.}\ \bibnamefont
  {Esch}}, \bibinfo {author} {\bibfnamefont {L.~V.}\ \bibnamefont {Bockstal}},
  \bibinfo {author} {\bibfnamefont {J.~D.}\ \bibnamefont {Boeck}}, \bibinfo
  {author} {\bibfnamefont {G.}~\bibnamefont {Verbanck}}, \bibinfo {author}
  {\bibfnamefont {A.~S.}\ \bibnamefont {van Steenbergen}}, \bibinfo {author}
  {\bibfnamefont {P.~J.}\ \bibnamefont {Wellmann}}, \bibinfo {author}
  {\bibfnamefont {B.}~\bibnamefont {Grietens}}, \bibinfo {author}
  {\bibfnamefont {R.}~\bibnamefont {Bogaerts}}, \bibinfo {author}
  {\bibfnamefont {F.}~\bibnamefont {Herlach}}, \ and\ \bibinfo {author}
  {\bibfnamefont {G.}~\bibnamefont {Borghs}},\ }\href@noop {} {\bibfield
  {journal} {\bibinfo  {journal} {Phys. Rev. B}\ }\textbf {\bibinfo {volume}
  {56}},\ \bibinfo {pages} {13103} (\bibinfo {year} {1997})}\BibitemShut
  {NoStop}%
\bibitem [{\citenamefont {Sinnecker}\ \emph {et~al.}(2010)\citenamefont
  {Sinnecker}, \citenamefont {Penello}, \citenamefont {Rappoport},
  \citenamefont {Sant'Anna}, \citenamefont {Souza}, \citenamefont {Pires},
  \citenamefont {Furdyna},\ and\ \citenamefont {Liu}}]{Sinnecker2010}%
  \BibitemOpen
  \bibfield  {author} {\bibinfo {author} {\bibfnamefont {E.~H. C.~P.}\
  \bibnamefont {Sinnecker}}, \bibinfo {author} {\bibfnamefont {G.~M.}\
  \bibnamefont {Penello}}, \bibinfo {author} {\bibfnamefont {T.~G.}\
  \bibnamefont {Rappoport}}, \bibinfo {author} {\bibfnamefont {M.~M.}\
  \bibnamefont {Sant'Anna}}, \bibinfo {author} {\bibfnamefont {D.~E.~R.}\
  \bibnamefont {Souza}}, \bibinfo {author} {\bibfnamefont {M.~P.}\ \bibnamefont
  {Pires}}, \bibinfo {author} {\bibfnamefont {J.~K.}\ \bibnamefont {Furdyna}},
  \ and\ \bibinfo {author} {\bibfnamefont {X.}~\bibnamefont {Liu}},\
  }\href@noop {} {\bibfield  {journal} {\bibinfo  {journal} {Phys. Rev. B}\
  }\textbf {\bibinfo {volume} {81}},\ \bibinfo {pages} {245203} (\bibinfo
  {year} {2010})}\BibitemShut {NoStop}%
\bibitem [{\citenamefont {Wang}\ \emph {et~al.}(2008)\citenamefont {Wang},
  \citenamefont {Campion}, \citenamefont {Rushforth}, \citenamefont {Edmonds},
  \citenamefont {Foxon},\ and\ \citenamefont {Gallagher}}]{Wang2008}%
  \BibitemOpen
  \bibfield  {author} {\bibinfo {author} {\bibfnamefont {M.}~\bibnamefont
  {Wang}}, \bibinfo {author} {\bibfnamefont {R.~P.}\ \bibnamefont {Campion}},
  \bibinfo {author} {\bibfnamefont {A.~W.}\ \bibnamefont {Rushforth}}, \bibinfo
  {author} {\bibfnamefont {K.~W.}\ \bibnamefont {Edmonds}}, \bibinfo {author}
  {\bibfnamefont {C.~T.}\ \bibnamefont {Foxon}}, \ and\ \bibinfo {author}
  {\bibfnamefont {B.~L.}\ \bibnamefont {Gallagher}},\ }\href@noop {} {\bibfield
   {journal} {\bibinfo  {journal} {Applied Physics Letters}\ }\textbf {\bibinfo
  {volume} {93}},\ \bibinfo {pages} {132103} (\bibinfo {year}
  {2008})}\BibitemShut {NoStop}%
\bibitem [{\citenamefont {Ziegler}\ \emph {et~al.}(1985)\citenamefont
  {Ziegler}, \citenamefont {Biersack},\ and\ \citenamefont
  {Littmark}}]{Ziegler1985}%
  \BibitemOpen
  \bibfield  {author} {\bibinfo {author} {\bibfnamefont {J.}~\bibnamefont
  {Ziegler}}, \bibinfo {author} {\bibfnamefont {J.}~\bibnamefont {Biersack}}, \
  and\ \bibinfo {author} {\bibfnamefont {U.}~\bibnamefont {Littmark}},\
  }\href@noop {} {\emph {\bibinfo {title} {The Stopping and Range of Ions in
  Solids}}}\ (\bibinfo  {publisher} {Pergamon Press},\ \bibinfo {address} {New
  York},\ \bibinfo {year} {1985})\BibitemShut {NoStop}%
\bibitem [{\citenamefont {Pons}\ and\ \citenamefont
  {Bourgoin}(1985)}]{Pons1985}%
  \BibitemOpen
  \bibfield  {author} {\bibinfo {author} {\bibfnamefont {D.}~\bibnamefont
  {Pons}}\ and\ \bibinfo {author} {\bibfnamefont {J.~C.}\ \bibnamefont
  {Bourgoin}},\ }\href@noop {} {\bibfield  {journal} {\bibinfo  {journal}
  {Journal of Physics C: Solid State Physics}\ }\textbf {\bibinfo {volume}
  {18}},\ \bibinfo {pages} {3839} (\bibinfo {year} {1985})}\BibitemShut
  {NoStop}%
\bibitem [{\citenamefont {Yuan}\ \emph {et~al.}(2016)\citenamefont {Yuan},
  \citenamefont {H\"{u}bner}, \citenamefont {Liu}, \citenamefont {Sawicki},
  \citenamefont {Gordan}, \citenamefont {Salvan}, \citenamefont {Zahn},
  \citenamefont {Banerjee}, \citenamefont {Baehtz}, \citenamefont {Helm},\ and\
  \citenamefont {Zhou}}]{Yuan2016}%
  \BibitemOpen
  \bibfield  {author} {\bibinfo {author} {\bibfnamefont {Y.}~\bibnamefont
  {Yuan}}, \bibinfo {author} {\bibfnamefont {R.}~\bibnamefont {H\"{u}bner}},
  \bibinfo {author} {\bibfnamefont {F.}~\bibnamefont {Liu}}, \bibinfo {author}
  {\bibfnamefont {M.}~\bibnamefont {Sawicki}}, \bibinfo {author} {\bibfnamefont
  {O.}~\bibnamefont {Gordan}}, \bibinfo {author} {\bibfnamefont
  {G.}~\bibnamefont {Salvan}}, \bibinfo {author} {\bibfnamefont {D.~R.~T.}\
  \bibnamefont {Zahn}}, \bibinfo {author} {\bibfnamefont {D.}~\bibnamefont
  {Banerjee}}, \bibinfo {author} {\bibfnamefont {C.}~\bibnamefont {Baehtz}},
  \bibinfo {author} {\bibfnamefont {M.}~\bibnamefont {Helm}}, \ and\ \bibinfo
  {author} {\bibfnamefont {S.}~\bibnamefont {Zhou}},\ }\href@noop {} {\bibfield
   {journal} {\bibinfo  {journal} {{ACS} Applied Materials {\&} Interfaces}\
  }\textbf {\bibinfo {volume} {8}},\ \bibinfo {pages} {3912} (\bibinfo {year}
  {2016})}\BibitemShut {NoStop}%
\bibitem [{\citenamefont {Cho}\ \emph {et~al.}(2007)\citenamefont {Cho},
  \citenamefont {Scarpulla}, \citenamefont {Liu}, \citenamefont {Zhou},
  \citenamefont {Dubon},\ and\ \citenamefont {Furdyna}}]{Cho2007}%
  \BibitemOpen
  \bibfield  {author} {\bibinfo {author} {\bibfnamefont {Y.~J.}\ \bibnamefont
  {Cho}}, \bibinfo {author} {\bibfnamefont {M.~A.}\ \bibnamefont {Scarpulla}},
  \bibinfo {author} {\bibfnamefont {X.}~\bibnamefont {Liu}}, \bibinfo {author}
  {\bibfnamefont {Y.~Y.}\ \bibnamefont {Zhou}}, \bibinfo {author}
  {\bibfnamefont {O.~D.}\ \bibnamefont {Dubon}}, \ and\ \bibinfo {author}
  {\bibfnamefont {J.~K.}\ \bibnamefont {Furdyna}},\ }in\ \href@noop {} {\emph
  {\bibinfo {booktitle} {{AIP} Conference Proceedings}}}\ (\bibinfo
  {publisher} {{AIP}},\ \bibinfo {year} {2007})\BibitemShut {NoStop}%
\bibitem [{\citenamefont {Hooge}(1969)}]{Hooge1969}%
  \BibitemOpen
  \bibfield  {author} {\bibinfo {author} {\bibfnamefont {F.}~\bibnamefont
  {Hooge}},\ }\href@noop {} {\bibfield  {journal} {\bibinfo  {journal} {Physics
  Letters A}\ }\textbf {\bibinfo {volume} {29}},\ \bibinfo {pages} {139}
  (\bibinfo {year} {1969})}\BibitemShut {NoStop}%
\bibitem [{\citenamefont {Hooge}(1976)}]{Hooge1976}%
  \BibitemOpen
  \bibfield  {author} {\bibinfo {author} {\bibfnamefont {F.}~\bibnamefont
  {Hooge}},\ }\href@noop {} {\bibfield  {journal} {\bibinfo  {journal} {Physica
  B}\ }\textbf {\bibinfo {volume} {83}},\ \bibinfo {pages} {14} (\bibinfo
  {year} {1976})}\BibitemShut {NoStop}%
\bibitem [{\citenamefont {Scofield}(1987)}]{Scofield1987}%
  \BibitemOpen
  \bibfield  {author} {\bibinfo {author} {\bibfnamefont {J.~H.}\ \bibnamefont
  {Scofield}},\ }\href@noop {} {\bibfield  {journal} {\bibinfo  {journal} {Rev.
  Sci. Instrum.}\ }\textbf {\bibinfo {volume} {58}},\ \bibinfo {pages} {985}
  (\bibinfo {year} {1987})}\BibitemShut {NoStop}%
\bibitem [{\citenamefont {Raquet}(2001)}]{Raquet2001}%
  \BibitemOpen
  \bibfield  {author} {\bibinfo {author} {\bibfnamefont {B.}~\bibnamefont
  {Raquet}},\ }in\ \href@noop {} {\emph {\bibinfo {booktitle} {{Spin
  Electronics, Chapter: "Electronic Noise in Magnetic Materials and
  Devices"}}}}\ (\bibinfo  {publisher} {Springer, Heidelberg},\ \bibinfo {year}
  {2001})\ pp.\ \bibinfo {pages} {232--273}\BibitemShut {NoStop}%
\bibitem [{\citenamefont {M\"{u}ller}(2011)}]{Mueller2011}%
  \BibitemOpen
  \bibfield  {author} {\bibinfo {author} {\bibfnamefont {J.}~\bibnamefont
  {M\"{u}ller}},\ }\href@noop {} {\bibfield  {journal} {\bibinfo  {journal}
  {{ChemPhysChem}}\ }\textbf {\bibinfo {volume} {12}},\ \bibinfo {pages} {1222}
  (\bibinfo {year} {2011})}\BibitemShut {NoStop}%
\bibitem [{\citenamefont {Matsukura}\ \emph {et~al.}(1998)\citenamefont
  {Matsukura}, \citenamefont {Ohno}, \citenamefont {Shen},\ and\ \citenamefont
  {Sugawara}}]{Matsukura1998}%
  \BibitemOpen
  \bibfield  {author} {\bibinfo {author} {\bibfnamefont {F.}~\bibnamefont
  {Matsukura}}, \bibinfo {author} {\bibfnamefont {H.}~\bibnamefont {Ohno}},
  \bibinfo {author} {\bibfnamefont {A.}~\bibnamefont {Shen}}, \ and\ \bibinfo
  {author} {\bibfnamefont {Y.}~\bibnamefont {Sugawara}},\ }\href@noop {}
  {\bibfield  {journal} {\bibinfo  {journal} {Phys. Rev. B}\ }\textbf {\bibinfo
  {volume} {57}},\ \bibinfo {pages} {R2037} (\bibinfo {year}
  {1998})}\BibitemShut {NoStop}%
\bibitem [{\citenamefont {Edmonds}\ \emph {et~al.}(2004)\citenamefont
  {Edmonds}, \citenamefont {Bogus{\l}awski}, \citenamefont {Wang},
  \citenamefont {Campion}, \citenamefont {Novikov}, \citenamefont {Farley},
  \citenamefont {Gallagher}, \citenamefont {Foxon}, \citenamefont {Sawicki},
  \citenamefont {Dietl}, \citenamefont {Nardelli},\ and\ \citenamefont
  {Bernholc}}]{Edmonds2004}%
  \BibitemOpen
  \bibfield  {author} {\bibinfo {author} {\bibfnamefont {K.~W.}\ \bibnamefont
  {Edmonds}}, \bibinfo {author} {\bibfnamefont {P.}~\bibnamefont
  {Bogus{\l}awski}}, \bibinfo {author} {\bibfnamefont {K.~Y.}\ \bibnamefont
  {Wang}}, \bibinfo {author} {\bibfnamefont {R.~P.}\ \bibnamefont {Campion}},
  \bibinfo {author} {\bibfnamefont {S.~N.}\ \bibnamefont {Novikov}}, \bibinfo
  {author} {\bibfnamefont {N.~R.~S.}\ \bibnamefont {Farley}}, \bibinfo {author}
  {\bibfnamefont {B.~L.}\ \bibnamefont {Gallagher}}, \bibinfo {author}
  {\bibfnamefont {C.~T.}\ \bibnamefont {Foxon}}, \bibinfo {author}
  {\bibfnamefont {M.}~\bibnamefont {Sawicki}}, \bibinfo {author} {\bibfnamefont
  {T.}~\bibnamefont {Dietl}}, \bibinfo {author} {\bibfnamefont {M.~B.}\
  \bibnamefont {Nardelli}}, \ and\ \bibinfo {author} {\bibfnamefont
  {J.}~\bibnamefont {Bernholc}},\ }\href@noop {} {\bibfield  {journal}
  {\bibinfo  {journal} {Phys. Rev. Lett.}\ }\textbf {\bibinfo {volume} {92}},\
  \bibinfo {pages} {037201} (\bibinfo {year} {2004})}\BibitemShut {NoStop}%
\bibitem [{\citenamefont {Lonsky}\ \emph {et~al.}(2017)\citenamefont {Lonsky},
  \citenamefont {Teschabai-Oglu}, \citenamefont {Pierz}, \citenamefont
  {Sievers}, \citenamefont {Schumacher}, \citenamefont {Yuan}, \citenamefont
  {B\"{o}ttger}, \citenamefont {Zhou},\ and\ \citenamefont
  {M\"{u}ller}}]{Lonsky2017}%
  \BibitemOpen
  \bibfield  {author} {\bibinfo {author} {\bibfnamefont {M.}~\bibnamefont
  {Lonsky}}, \bibinfo {author} {\bibfnamefont {J.}~\bibnamefont
  {Teschabai-Oglu}}, \bibinfo {author} {\bibfnamefont {K.}~\bibnamefont
  {Pierz}}, \bibinfo {author} {\bibfnamefont {S.}~\bibnamefont {Sievers}},
  \bibinfo {author} {\bibfnamefont {H.~W.}\ \bibnamefont {Schumacher}},
  \bibinfo {author} {\bibfnamefont {Y.}~\bibnamefont {Yuan}}, \bibinfo {author}
  {\bibfnamefont {R.}~\bibnamefont {B\"{o}ttger}}, \bibinfo {author}
  {\bibfnamefont {S.}~\bibnamefont {Zhou}}, \ and\ \bibinfo {author}
  {\bibfnamefont {J.}~\bibnamefont {M\"{u}ller}},\ }\href@noop {} {\bibfield
  {journal} {\bibinfo  {journal} {Acta Physica Polonica A (accepted)}\ }
  (\bibinfo {year} {2017})}\BibitemShut {NoStop}%
\bibitem [{\citenamefont {Dutta}\ \emph {et~al.}(1979)\citenamefont {Dutta},
  \citenamefont {Dimon},\ and\ \citenamefont {Horn}}]{Dutta1979}%
  \BibitemOpen
  \bibfield  {author} {\bibinfo {author} {\bibfnamefont {P.}~\bibnamefont
  {Dutta}}, \bibinfo {author} {\bibfnamefont {P.}~\bibnamefont {Dimon}}, \ and\
  \bibinfo {author} {\bibfnamefont {P.~M.}\ \bibnamefont {Horn}},\ }\href@noop
  {} {\bibfield  {journal} {\bibinfo  {journal} {Phys. Rev. Lett.}\ }\textbf
  {\bibinfo {volume} {43}},\ \bibinfo {pages} {646} (\bibinfo {year}
  {1979})}\BibitemShut {NoStop}%
\bibitem [{\citenamefont {Edmonds}\ \emph {et~al.}(2002)\citenamefont
  {Edmonds}, \citenamefont {Wang}, \citenamefont {Campion}, \citenamefont
  {Neumann}, \citenamefont {Foxon}, \citenamefont {Gallagher},\ and\
  \citenamefont {Main}}]{Edmonds2002}%
  \BibitemOpen
  \bibfield  {author} {\bibinfo {author} {\bibfnamefont {K.~W.}\ \bibnamefont
  {Edmonds}}, \bibinfo {author} {\bibfnamefont {K.~Y.}\ \bibnamefont {Wang}},
  \bibinfo {author} {\bibfnamefont {R.~P.}\ \bibnamefont {Campion}}, \bibinfo
  {author} {\bibfnamefont {A.~C.}\ \bibnamefont {Neumann}}, \bibinfo {author}
  {\bibfnamefont {C.~T.}\ \bibnamefont {Foxon}}, \bibinfo {author}
  {\bibfnamefont {B.~L.}\ \bibnamefont {Gallagher}}, \ and\ \bibinfo {author}
  {\bibfnamefont {P.~C.}\ \bibnamefont {Main}},\ }\href@noop {} {\bibfield
  {journal} {\bibinfo  {journal} {Appl. Phys. Lett.}\ }\textbf {\bibinfo
  {volume} {81}},\ \bibinfo {pages} {3010} (\bibinfo {year}
  {2002})}\BibitemShut {NoStop}%
\bibitem [{\citenamefont {Kogan}(1996)}]{Kogan1996}%
  \BibitemOpen
  \bibfield  {author} {\bibinfo {author} {\bibfnamefont {S.}~\bibnamefont
  {Kogan}},\ }\href@noop {} {\emph {\bibinfo {title} {{Electronic Noise and
  Fluctuations in Solids}}}}\ (\bibinfo  {publisher} {Cambrigde University
  Press},\ \bibinfo {year} {1996})\BibitemShut {NoStop}%
\bibitem [{\citenamefont {Rammal}\ \emph {et~al.}(1985)\citenamefont {Rammal},
  \citenamefont {Tannous}, \citenamefont {Breton},\ and\ \citenamefont
  {Tremblay}}]{Rammal1985}%
  \BibitemOpen
  \bibfield  {author} {\bibinfo {author} {\bibfnamefont {R.}~\bibnamefont
  {Rammal}}, \bibinfo {author} {\bibfnamefont {C.}~\bibnamefont {Tannous}},
  \bibinfo {author} {\bibfnamefont {P.}~\bibnamefont {Breton}}, \ and\ \bibinfo
  {author} {\bibfnamefont {A.~M.~S.}\ \bibnamefont {Tremblay}},\ }\href@noop {}
  {\bibfield  {journal} {\bibinfo  {journal} {Phys. Rev. Lett.}\ }\textbf
  {\bibinfo {volume} {54}},\ \bibinfo {pages} {1718} (\bibinfo {year}
  {1985})}\BibitemShut {NoStop}%
\bibitem [{\citenamefont {Yagil}\ and\ \citenamefont
  {Deutscher}(1992)}]{Yagil1992}%
  \BibitemOpen
  \bibfield  {author} {\bibinfo {author} {\bibfnamefont {Y.}~\bibnamefont
  {Yagil}}\ and\ \bibinfo {author} {\bibfnamefont {G.}~\bibnamefont
  {Deutscher}},\ }\href@noop {} {\bibfield  {journal} {\bibinfo  {journal}
  {Phys. Rev. B}\ }\textbf {\bibinfo {volume} {46}},\ \bibinfo {pages} {16115}
  (\bibinfo {year} {1992})}\BibitemShut {NoStop}%
\bibitem [{\citenamefont {Bae}\ and\ \citenamefont {Raebiger}(2016)}]{Bae2016}%
  \BibitemOpen
  \bibfield  {author} {\bibinfo {author} {\bibfnamefont {S.}~\bibnamefont
  {Bae}}\ and\ \bibinfo {author} {\bibfnamefont {H.}~\bibnamefont {Raebiger}},\
  }\href@noop {} {\bibfield  {journal} {\bibinfo  {journal} {Phys. Rev. B}\
  }\textbf {\bibinfo {volume} {94}},\ \bibinfo {pages} {241115} (\bibinfo
  {year} {2016})}\BibitemShut {NoStop}%
\bibitem [{\citenamefont {Khalid}\ \emph {et~al.}(2014)\citenamefont {Khalid},
  \citenamefont {Weschke}, \citenamefont {Skorupa}, \citenamefont {Helm},\ and\
  \citenamefont {Zhou}}]{Khalid2014}%
  \BibitemOpen
  \bibfield  {author} {\bibinfo {author} {\bibfnamefont {M.}~\bibnamefont
  {Khalid}}, \bibinfo {author} {\bibfnamefont {E.}~\bibnamefont {Weschke}},
  \bibinfo {author} {\bibfnamefont {W.}~\bibnamefont {Skorupa}}, \bibinfo
  {author} {\bibfnamefont {M.}~\bibnamefont {Helm}}, \ and\ \bibinfo {author}
  {\bibfnamefont {S.}~\bibnamefont {Zhou}},\ }\href@noop {} {\bibfield
  {journal} {\bibinfo  {journal} {Phys. Rev. B}\ }\textbf {\bibinfo {volume}
  {89}},\ \bibinfo {pages} {121301} (\bibinfo {year} {2014})}\BibitemShut
  {NoStop}%
\bibitem [{\citenamefont {Park}(2002)}]{Park2002}%
  \BibitemOpen
  \bibfield  {author} {\bibinfo {author} {\bibfnamefont {Y.~D.}\ \bibnamefont
  {Park}},\ }\href@noop {} {\bibfield  {journal} {\bibinfo  {journal}
  {Science}\ }\textbf {\bibinfo {volume} {295}},\ \bibinfo {pages} {651}
  (\bibinfo {year} {2002})}\BibitemShut {NoStop}%
\end{thebibliography}%

%%%%%%%%%%%%%%%%%%%%%%%%%%%%%%%%%%%%%%%%%%%%%%%%%%%%%%%%%%%%%%%%%%%%%%%%%%%%%%%%%%%%%%%%%%%%%%%%%%%%%%%%%%%%%%%%%%%%%%%%%%%%%%%%%%%

\clearpage

\end{document}